\documentclass[10pt,preprint]{aastex}
\usepackage{times}
\newcommand{\vect}[1]{\ensuremath{\mbox{\boldmath $#1$}}}
\newcommand{\eb}{EROS~BLG-2000-5}
\newcommand{\mb}{MACHO~97-BLG-41}

\newcommand{\RHOSTAR}{4.80\pm 0.04}
\newcommand{\THETAE}{1.38\pm 0.12\ \mbox{mas}}
\newcommand{\MASS}{0.612\pm 0.057\ \mbox{M}_\sun}
\newcommand{\RETILDE}{3.61\pm 0.11\ \mbox{AU}}
\newcommand{\PIE}{0.277\pm 0.008}
\newcommand{\TE}{76.8\pm 2.1\ \mbox{days}}
\newcommand{\MUEANEC}{36\fdg3}
\newcommand{\MUEANGL}{23\fdg9}
\newcommand{\VTILDEE}{-74.5\pm 3.1}
\newcommand{\VTILDEN}{33.0\pm 11.0}
\newcommand{\DREL}{2.62\pm 0.24\ \mbox{kpc}}
\newcommand{\PIREL}{0.382\pm 0.035\ \mbox{mas}}
\newcommand{\MURELA}{18.0\pm 1.7\ \mbox{$\mu$as day$^{-1}$}}
\newcommand{\MURELB}{31.1\pm 2.9\ \mbox{km s$^{-1}$ kpc$^{-1}$}}

\newcommand{\EPST}{25.9\pm 5.1}

\newcommand{\PROJLT}{0.0013\pm 0.0341\ \mbox{AU}}

\newcommand{\AMINT}{2.87\pm 0.09\ \mbox{AU}}

\newcommand{\ECCMINT}{0.021\pm 0.281}

\slugcomment
{to appear in the Astrophysical Journal (v572n1; 2002 June 10)}
\shorttitle{FIRST MICROLENS MASS MEASUREMENT}
\shortauthors{the PLANET collaboration}

\begin{document}

\title{
First microlens mass measurement:\\
PLANET photometry of \eb}

\author{
Jin H. An\altaffilmark{1},
M. D. Albrow\altaffilmark{2,3},
J.-P. Beaulieu\altaffilmark{4},
J. A. R. Caldwell\altaffilmark{5},
D. L. DePoy\altaffilmark{1},\\
M. Dominik\altaffilmark{6},
B. S. Gaudi\altaffilmark{7,8},
A. Gould\altaffilmark{1},
J. Greenhill\altaffilmark{9},
K. Hill\altaffilmark{9},
S. Kane\altaffilmark{9,10},\\
R. Martin\altaffilmark{11},
J. Menzies\altaffilmark{5},
R. W. Pogge\altaffilmark{1},
K. R. Pollard\altaffilmark{2,12},
P. D. Sackett\altaffilmark{6},\\
K. C. Sahu\altaffilmark{3},
P. Vermaak\altaffilmark{5},
R. Watson\altaffilmark{9},
and
A. Williams\altaffilmark{11}
}\author{(The PLANET Collaboration)}

\altaffiltext{1}
{Department of Astronomy, the Ohio State University,
140 West 18th Avenue, Columbus, OH 43210, USA}
\altaffiltext{2}
{Department of Physics \& Astronomy, University of Canterbury,
Private Bag 4800, Christchurch, New Zealand}
\altaffiltext{3}
{Space Telescope Science Institute,
3700 San Martin Drive, Baltimore, MD 21218, USA}
\altaffiltext{4}
{Institut d'Astrophysique de Paris, INSU-CNRS,
98 bis Boulevard Arago, F 75014 Paris, France}
\altaffiltext{5}
{South African Astronomical Observatory,
P.O. Box 9, Observatory, 7935 South Africa}
\altaffiltext{6}
{Kapteyn Institute, Rijksuniversiteit Groningen,
Postbus 800, 9700 AV Groningen, The Netherlands}
\altaffiltext{7}
{School of Natural Sciences, Institute for Advanced Study,
Einstein Drive, Princeton, NJ 08540, USA}
\altaffiltext{8}{Hubble Fellow}
\altaffiltext{9}
{Physics Department, University of Tasmania,
G.P.O. 252C, Hobart, Tasmania 7001, Australia}
\altaffiltext{10}
{School of Physics \& Astronomy, University of St.~Andrews,
North Haugh, St.~Andrews, Fife KY16 9SS, UK}
\altaffiltext{11}
{Perth Observatory,
Walnut Road, Bickley, Western Australia 6076, Australia}
\altaffiltext{12}
{Physics Department, Gettysburg College,
300 North Washington Street, Gettysburg, PA 17325, USA}

\begin{abstract}
We analyze PLANET photometric observations of the caustic-crossing
binary-lens microlensing event, \eb, and find that modeling the
observed light curve requires incorporation of the microlens
parallax and the binary orbital motion. The projected Einstein radius
($\tilde r_{\rm E} = \RETILDE$) is derived from the measurement of
the microlens parallax, and we are also able to infer the angular
Einstein radius ($\theta_{\rm E} = \THETAE$) from the finite source
effect on the light curve, combined with an estimate of the angular
size of the source given by the source position in a color-magnitude
diagram. The lens mass, $M = \MASS$, is found
by combining these two quantities.
This is the first time that parallax effects are detected for a
caustic-crossing event and also the first time that the lens mass degeneracy
has been completely broken through photometric monitoring alone.
The combination of $\tilde r_{\rm E}$ and $\theta_{\rm E}$
also allows us to conclude that the lens
lies in the near side of the disk, within $2.6\ \mbox{kpc}$ of the Sun,
while the radial velocity measurement indicates that the source is
a Galactic bulge giant.

\end{abstract}
\keywords{gravitational lensing --- binaries: general}

\section{Introduction
\label{sec:intro}}

Although an initial objective of microlensing experiments was to probe
the mass distribution of compact sub-luminous objects in the
Galactic halo \citep{Pa86}, the determination of the individual
lens masses was in general believed not to be possible because the only
generically measurable quantity is a degenerate combination of the mass with
lens-source relative parallax and proper motion. However, \citet{Go92} showed
that there are additional observables that in principle can
be used to break the degeneracy between physical parameters
and so yield a measurement of the lens mass. If one independently
measures the angular Einstein radius, $\theta_{\rm E}$, and the
projected Einstein radius, $\tilde r_{\rm E}$:
\begin{equation}
\label{eqn:thetare}
\theta_{\rm E} \equiv \left(\frac{2 R_{\rm sch}}{D_{\rm rel}}\right)^{1/2}
\,;\ \ \
\tilde r_{\rm E} \equiv \left(2 R_{\rm sch} D_{\rm rel}\right)^{1/2}
\,,\end{equation}
then the lens mass $M$ is obtained by
\begin{eqnarray}
\label{eqn:lensmass}
M&=&\frac{c^2}{4 G} \tilde r_{\rm E} \theta_{\rm E}
\nonumber\\&=&0.1227\ \mbox{M$_\sun$}
\left(\frac{\tilde r_{\rm E}}{1\ \mbox{AU}}\right)
\left(\frac{\theta_{\rm E}}{1\ \mbox{mas}}\right)
\ .\end{eqnarray}
Here $R_{\rm sch} \equiv 2 G M c^{-2}$ is the Schwarzschild
radius of the lens mass,
$D_{\rm rel}^{-1} = D_{\rm L}^{-1} - D_{\rm S}^{-1}$, and $D_{\rm L}$ 
and $D_{\rm S}$ are the distances to the lens and the source.

Measurement of $\theta_{\rm E}$ requires a standard ruler on the plane
of the sky to be compared with the Einstein ring. For the subset of
microlensing events in which the source passes very close to or directly
over a caustic (region of singularity of the lens mapping at which
the magnification for a point source is formally infinite), the finite
source size affects the light curve. The source
radius in units of the Einstein radius, $\rho_\ast$, can then be
determined through analysis of the light curve, and thus, one
can determine $\theta_{\rm E}$ ($= \theta_\ast/\rho_\ast$) once the
angular radius of the source, $\theta_\ast$, is estimated from its
de-reddened color and apparent magnitude. Although this idea was originally
proposed for point-mass lenses, which have point-like caustics
\citep{Go94, NW94, WM94}, in practice it has been used mainly for
binary lenses, which have line-like caustics and hence much larger
cross sections 
\citep{MA02, MA03, PL_SMC, PL_MB9741, PL_OB9923, SMC}.

All ideas proposed to measure $\tilde r_{\rm E}$ are based on
the detection of parallax effects 
\citep*{Re66, GKR86, Go92, Go95, GMB94, HW95, HW96, Ho99, AG01}:
either
measuring the difference in the event as observed simultaneously
from two or more locations, or
observing the source from a frame that 
accelerates substantially during the course of the event.
The prime example of the latter
is the Earth's orbital motion (annual parallax).
For these cases, it is
convenient to re-express $\tilde r_{\rm E}$ as the microlens parallax,
${\pi}_{\rm E}$,
\begin{equation}
\label{eqn:pie}
\pi_{\rm E} \equiv \frac{\pi_{\rm L} - \pi_{\rm S}}{\theta_{\rm E}} =
\frac{\mbox{1-kpc}/D_{\rm rel}}{\theta_{\rm E}/\mbox{1-mas}} =
\frac{1\ \mbox{AU}}{\tilde r_{\rm E}}
\ .\end{equation}
Here, $\pi_{\rm L}$ and $\pi_{\rm S}$ are 
the (annual) trigonometric parallax of the
lens and the source so that $\pi_{\rm X}/\mbox{1-mas} = \mbox{1-kpc}/D_{\rm X}$,
where subscript `X' is either `L'(ens) or `S'(ource). 
To date, $\pi_{\rm E}$ has been measured
for several events by detecting the distortion
in the light curve due to the annual parallax effect
\citep*{MA01, Ma99, OGLE, MOA, OGLEII, O9932, MPS_PAR}

\citet{AG01} argue that $\pi_{\rm E}$ is measurable for a caustic-crossing
binary event exhibiting a well-observed peak caused by a
cusp approach in addition to the two usual caustic crossings.
Hence, one can determine the lens mass for such an event, 
since $\theta_{\rm E}$ can be estimated from any well-observed
caustic crossing. The caustic-crossing binary event \eb\ is
archetypal of such events. In addition, the event has a relatively
long time scale ($t_{\rm E} \sim 100\ \mbox{days}$), 
which is generally favorable
for the measurement of $\pi_{\rm E}$.

In fact, \eb\ features many unique characteristics not only
in the intrinsic nature of the event but also in the observations of it.
These include a moderately well-covered first caustic crossing
(entrance), a timely prediction -- of not only the timing 
but also the duration -- of the second caustic crossing
(exit), the unprecedented four-day length of the second crossing,
and two time-series of spectral observations of the source during
the second crossing \citep{VLT, Keck}. To fully understand and utilize
this wealth of information, however, requires
detailed quantitative modeling of the event. Here, we present the
first model of \eb\ based on the photometric
observations made by the Probing Lensing Anomalies NETwork (PLANET)\footnote{
\url{http://www.astro.rug.nl/\~{}planet/}} collaboration.
In the current paper, we
focus on the geometry of the event
and find that both parallax and projected binary orbital motion
are required to
successfully model the light curve. Furthermore,
we also measure the angular Einstein radius from the finite source
effect during caustic crossings and the source angular size
derived from the source position in a color-magnitude diagram (CMD).
Hence, this is the first event for which both
$\tilde r_{\rm E}$ and $\theta_{\rm E}$ are measured simultaneously
and so for which the lens mass is unambiguously measured.

\section{Data
\label{sec:dat}}

\begin{figure}
\plotone{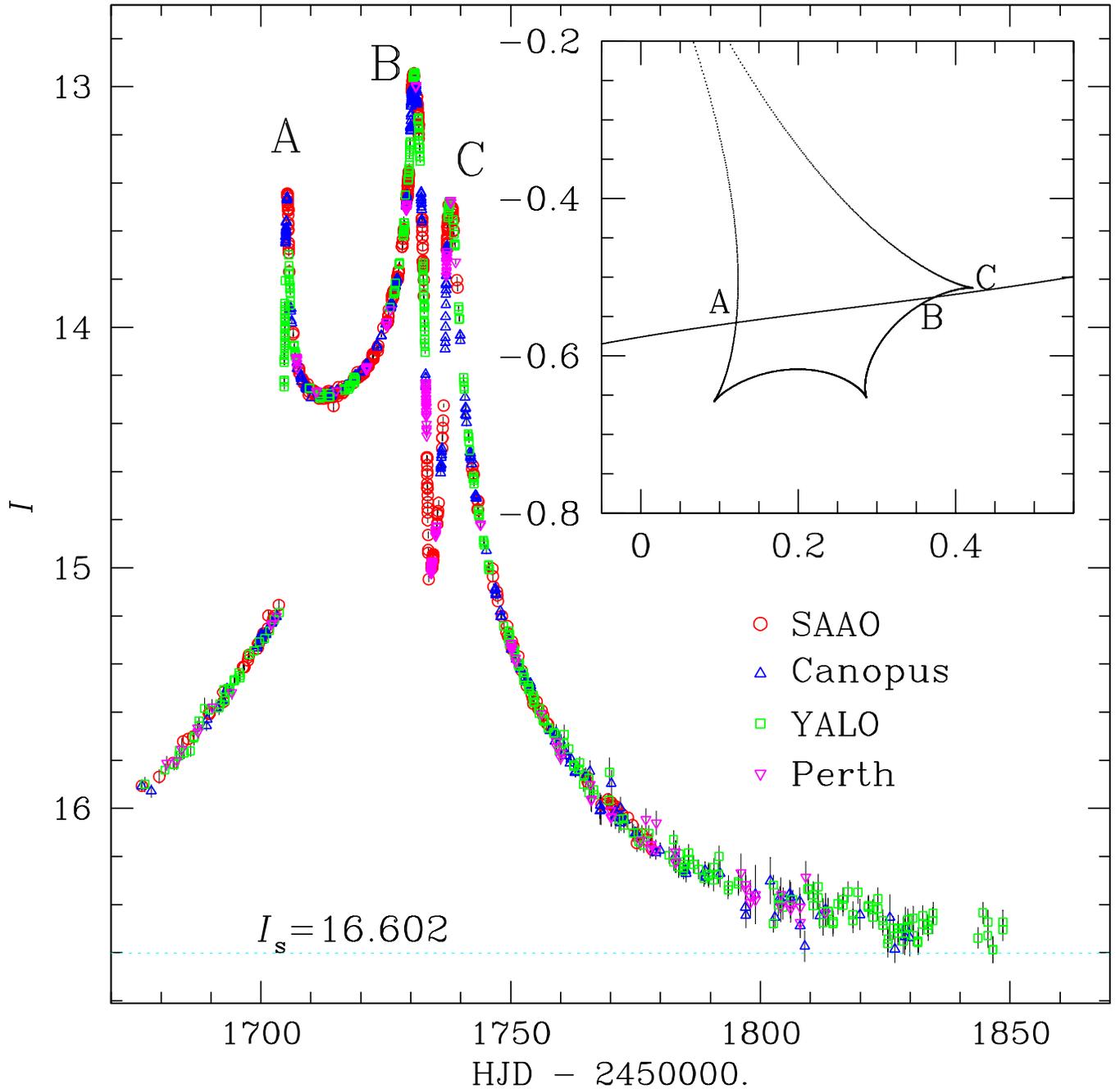}
\caption{\label{fig:ltc}
PLANET $I$-band light curve of \eb\ (the 2000 season only).
Only the data points used for the analysis (``cleaned high-quality'' subset;
see \S~\ref{sec:dat}) are plotted. Data shown are from SAAO ({\it red circles}),
Canopus ({\it blue triangles}), YALO ({\it green squares}),
and Perth ({\it magenta inverted triangles}).
All data points have been de-blended
using the fit result -- also accounting for the seeing correction --
and transformed to the standard $I$ magnitude;
$I=I_{\rm s} - 2.5\ \log[(F(t)-F_{\rm b})/F_{\rm s}]$.
The calibrated source magnitude ($I_{\rm s}=16.602$)
is shown as a dotted line.
The three bumps in the light curve and the corresponding positions
relative to the microlens geometry are also indicated.
}\end{figure}

On 2000 May 5, the Exp\'erience de Recherche
d'Objets Sombres (EROS)\footnote{
\url{http://www-dapnia.cea.fr/Spp/Experiences/EROS/alertes.html}}
collaboration issued an alert that \eb\ was a probable
microlensing event ($\mbox{RA} = 17^{\rm h} 53^{\rm m} 11\fs5$,
$\mbox{Dec} = -30\arcdeg 55\arcmin 35 \arcsec$;
$l = 359\fdg14$, $b = -2\fdg43$).
On 2000 June 8, the Microlensing Planet Search 
(MPS)\footnote{
\url{http://bustard.phys.nd.edu/MPS/}} collaboration
issued an anomaly alert, saying that the source had brightened by 0.5~mag
from the previous night and had continued to brighten by 0.1~mag in
40~minutes. PLANET intensified its observations immediately, and has
been continuing to monitor the event up to the present (2001 August).
Observations for PLANET were made from
four telescopes: the Canopus 1~m near Hobart, Tasmania, Australia;
the Perth/Lowell 0.6~m at Bickley, Western Australia, Australia;
the Elizabeth 1~m at the South African Astronomical Observatory (SAAO)
at Sutherland, South Africa; and the Yale/AURA/Lisbon/OSU (YALO) 1~m
at the Cerro Tololo Interamerican Observatory (CTIO) at La Serena, Chile.
Data were taken in $V$ (except Perth), $I$ (all four sites),
and $H$ (SAAO and YALO) bands. For the present study,
we make use primarily of the $I$ band data. That is, we fit for the model
using only $I$ band data, while the $V$ band data are used 
(together with the $I$ data) only to determine
the position of the source on CMD (see \S~\ref{sec:thetae}).
The light curve (Fig.~\ref{fig:ltc})
exhibits two peaks that have the characteristic forms of 
an entrance (A) and exit (B) caustic crossing
\citep*[][also see fig.~1 of \citealt{GA99}]{SEF92}
immediately followed by a third peak (C)
which is caused by the passage of the source close to a cusp.

The data have been reduced in a usual way and the photometry on them
has been performed by point spread function (PSF) fitting 
using DoPHOT \citep*{dophot}.
The relative photometric scaling between the different telescopes is
determined as part of the fit to the light curve which includes the independent
determinations of the source and the background fluxes at each telescope.
We find, as was the case for several previous events, that
due to the crowdedness of the field, the moderate seeing conditions,
and possibly some other unidentified systematics, the amount of blended
light entering the PSF is affected by seeing, and that the formal errors
reported by DoPHOT tend to underestimate the actual photometric uncertainties.
We tackle these problems by incorporating a linear seeing correction for the
background flux and rescaling the error bars to force the reduced $\chi^2$ 
of our best model to unity. For details of these procedures, see
\citet{PL_MB9741, PL_OB9814, PL_OB9923} and \citet{PL_FIVE}.

We also analyze the data by difference imaging, mostly using
ISIS \citep{Al00}. We compare the scatter of the
photometry on the difference images to that of the direct PSF
fit photometry by deriving the normalized-summed squares of the
signal-to-noise ratios:
\begin{equation}
\label{eqn:q}
{\mathcal{Q}} \equiv
\sum_i \left(\frac{F_{\rm s} A_i}{\tilde\sigma_i}\right)^2 =
\frac{N}{\chi^2}
\sum_i \left(\frac{F_{\rm s} A_i}{\sigma_i}\right)^2
\,,\end{equation}
where $F_{\rm s}$ is the source flux derived from the model,
$A_i$ is the magnification predicted by the model for the data
point, $\sigma_i$ and $\tilde\sigma_i$ are the photometric
uncertainty (in flux) for individual data before and after
rescaling the error bars [i.e.\ $\tilde\sigma_i=\sigma_i(\chi^2/N)^{1/2}$],
and $N$ is the number of the data points.
Strictly speaking, ${\mathcal{Q}}$ defined as in equation~(\ref{eqn:q})
is model-dependent, but
if the data sets to be compared do not differ with
one another in systematic ways (and the chosen model is close enough
to the real one), ${\mathcal{Q}}$ can be used as a proxy for the
relative statistical weight given by the data set without notable biases.
We find that difference imaging significantly improves the stability
of the photometry for the data from Canopus and Perth, but it
somewhat worsens the photometry for
the data from SAAO and YALO. We suspect that the result is related to
the overall seeing condition for the specific site, but a definite
conclusion will require more detailed study and would be beyond the scope
of the current paper. We hope to be able to further improve 
the photometry on difference images in the future, but for the
current analysis, we choose to use the data set with the better
${\mathcal{Q}}$ so that only for Canopus and Perth data sets,
we replace the result of the direct PSF photometry with
the difference imaging analysis result.

For the final analysis and the results reported here,
we have used only a ``high quality'' subset of the data.
Prior to any attempt
to model the event, we first exclude various faulty frames
and problematic photometry reported by the reduction/photometry
software. In addition, data points exhibiting large (formal)
errors and/or poor seeing compared to the rest of the data from
the same site are eliminated prior to the analysis. In particular,
the thresholds for the seeing cuts are chosen at the point where
the behavior of ``seeing-dependent background'' becomes noticeably
non-linear with the seeing variation. 
The criterion of the seeing and error cut for each
data set is reported in Table~\ref{tab:dat} together with
other photometric information.
In conjunction with the proper determination of the error rescaling
factors, we also remove isolated outliers as in \citet{PL_OB9923}.
Following these steps,
the ``cleaned high-quality'' $I$-band data set
consists of 1286 (= 403 SAAO + 333 Canopus + 389 YALO + 161 Perth)
measurements made during the 2000 season (between May 11 and November 12)
plus 60 additional observations (= 25 SAAO + 35 YALO) made in the 2001 season.
Finally, we exclude 49 data points (= 19 SAAO + 20 Canopus + 10 Perth)
that are very close to the cusp approach 
[$2451736.8 <$ Heliocentric Julian Date (HJD) $< 2451737.6$]
while we fit the light curve.
We find that the limited numerical resolution 
of the source, which in turn is dictated by computational considerations, 
introduces errors in the evaluation of $\chi^2$ in this region of the order
of a few, and in a way that does not smoothly depend on the parameters.
These would prevent us from finding the true minimum, or properly evaluating
the statistical errors. However, for the final model, we evaluate the
predicted fluxes and residuals for these points.
As we show in \S~\ref{sec:pie},
these resdiuals do not differ qualitatively from other residuals to the
fit.

\section{Parameterization
\label{sec:par}}

To specify the light curve of a static binary event 
with a rectilinear source trajectory
requires seven geometric parameters:
$d$, the binary separation in units of $\theta_{\rm E}$;
$q$, the binary mass ratio;
$\alpha$, the angle of the source-lens relative motion
with respect to the binary axis;
$t_{\rm E}$, the Einstein timescale
(the time required for the source to transit the Einstein radius);
$u_0$, the minimum angular separation
between the source and the binary center
-- either the geometric center or the center of mass --
in units of $\theta_{\rm E}$;
$t_0$, the time at this minimum;
$\rho_\ast$, the source size in units of $\theta_{\rm E}$.
(In addition, limb-darkening parameters for each wave band of observations,
and the source flux $F_{\rm s}$ and background flux $F_{\rm b}$
for each telescope and wavelength band are also required to transform
the light curve to a specific photometric system.)
Most generally, to incorporate the annual parallax and the projection of
binary orbital motion into the model,
one needs four additional parameters.
However, their inclusion,
especially of the parallax parameters, is not a trivial procedure,
since the natural coordinate basis for the description of the parallax
is the ecliptic system while the binary magnification pattern
possesses its own preferred direction, i.e.\ the binary axis. In the
following, we establish a consistent system to describe the
complete set of the eleven geometric parameters.

\subsection{Description of Geometry
\label{ssec:par}}

First, we focus on the description of parallax.
On the plane of the sky, the angular positions of the lens and the source
(seen from the center of the Earth)
are expressed generally by, 
\begin{mathletters}
\label{eqn:phi}
\begin{equation}
\vect{\varphi}_{\rm S} (t) = \vect{\varphi}_{{\rm S, c}}
+ (t - t_{\rm c}) \vect{\mu}_{\rm S} + \pi_{\rm S} \vect{\varsigma} (t)
\,,\end{equation}
\begin{equation}
\vect{\varphi}_{\rm L} (t) = \vect{\varphi}_{{\rm L, c}}
+ (t - t_{\rm c}) \vect{\mu}_{\rm L} + \pi_{\rm L} \vect{\varsigma} (t)
\ .\end{equation}
\end{mathletters}
Here $\vect{\varphi}_{{\rm S, c}}$ and $\vect{\varphi}_{{\rm L, c}}$
are the positions of the lens and the source at some reference time,
$t=t_{\rm c}$, as they would be observed from the Sun, $\vect{\mu}_{\rm S}$
and $\vect{\mu}_{\rm L}$ are the (heliocentric) proper motion of the lens
and the source, and
$\vect{\varsigma} (t)$ is the Sun's position vector with respect to the Earth,
projected onto the plane of the sky and normalized by an astronomical unit
(see Appendix~\ref{asec:sun}).
At any given time, $t$,
$\vect{\varsigma} (t)$ is completely determined with respect to an
ecliptic coordinate basis,
once the event's (ecliptic) coordinates are known.
For example, in the case of \eb\
($\lambda = 268\fdg53$, $\beta = -7\fdg50$),
$\vect{\varsigma} = (0, r_\earth \sin\beta)$ at approximately
2000 June 19, where 
$r_\earth$ is the distance between the Sun and the Earth at this
time in astronomical units ($r_\earth = 1.016$).
Then, the angular separation vector between the source and the lens
in units of $\theta_{\rm E}$ becomes
\begin{equation}
\label{eqn:udef}
\vect{u} (t) \equiv
\frac{\vect{\varphi}_{\rm S} - \vect{\varphi}_{\rm L}}{\theta_{\rm E}} =
\vect{\upsilon} + (t - t_{\rm c}) \vect{\mu}_{\rm E}
- \pi_{\rm E} \vect{\varsigma} (t)
\,,\end{equation}
where 
$\vect{\upsilon} \equiv
(\vect{\varphi}_{{\rm S, c}} - \vect{\varphi}_{{\rm L, c}})/\theta_{\rm E}$,
$\vect{\mu}_{\rm E} \equiv
(\vect{\mu}_{\rm S} - \vect{\mu}_{\rm L})/\theta_{\rm E}$,
and $\pi_{\rm E}$ is defined as in equation~(\ref{eqn:pie}).
Although equation~(\ref{eqn:udef}) is the most natural form of expression
for the parallax-affected trajectory, it is convenient to re-express
equation~(\ref{eqn:udef}) as the sum of the (geocentric) rectilinear motion at
the reference time and the parallactic deviations. In order to do this, we
evaluate $\vect{u}$ and $\dot{\vect{u}}$ at $t=t_{\rm c}$,
\begin{mathletters}
\label{eqn:attc}
\begin{equation}
\vect{u}_{t_{\rm c}} \equiv \vect{u} (t_{\rm c}) =
\vect{\upsilon} - \pi_{\rm E} \vect{\varsigma}_{t_{\rm c}}
\,,\end{equation}
\begin{equation}
\dot{\vect{u}}_{t_{\rm c}} \equiv \dot{\vect{u}} (t_{\rm c}) =
\vect{\mu}_{\rm E} - \pi_{\rm E} \dot{\vect{\varsigma}}_{t_{\rm c}}
\,,\end{equation}
\end{mathletters}
where $\vect{\varsigma}_{t_{\rm c}} \equiv \vect{\varsigma} (t_{\rm c})$ and
$\dot{\vect{\varsigma}}_{t_{\rm c}} \equiv \dot{\vect{\varsigma}} (t_{\rm c})$.
Solving equations~(\ref{eqn:attc}) for $\vect{\mu}_{\rm E}$ and
$\vect{\upsilon}$ and substituting them into equation~(\ref{eqn:udef}),
one obtains
\begin{equation}
\label{eqn:uform}
\vect{u} (t) = \vect{u}_{t_{\rm c}} + 
(t - t_{\rm c}) \dot{\vect{u}}_{t_{\rm c}} -
\pi_{\rm E} \vect{\mathcal{D}}_{\rm P}
\,,\end{equation}
where $\vect{\mathcal{D}}_{\rm P} \equiv
\vect{\varsigma} (t) - \vect{\varsigma}_{t_{\rm c}}
- (t - t_{\rm c}) \dot{\vect{\varsigma}}_{t_{\rm c}}$ is the
parallactic deviation.
Note that $\vect{\mathcal{D}}_{\rm P} 
\simeq (\ddot{\vect{\varsigma}}_{t_{\rm c}}/2)
(t-t_{\rm c})^{2}$ for $t \sim t_{\rm c}$, i.e.\
on relatively short time scales, the effect of the parallax
is equivalent to a uniform acceleration of 
$-\pi_{\rm E} \ddot{\vect{\varsigma}}_{t_{\rm c}}$. 
Equation~(\ref{eqn:uform}) is true in general for any microlensing
event including point-source/point-lens events. 

Next, we introduce the binary lens system. Whereas the parallax-affected
trajectory (eq.~[\ref{eqn:uform}]) is most naturally described in the
ecliptic coordinate system, the magnification pattern of the binary lens
is specified with respect to the binary axis. Hence, to construct a light
curve, one must transform the trajectory from ecliptic coordinates
to the binary coordinates. If the origins of both coordinates are chosen to
coincide at the binary center of mass, this transformation becomes purely
rotational. Thus, this basically adds one parameter to the problem:
the orientation of the binary axis in ecliptic coordinates.
In accordance with the parameterization of the projected binary orbital
motion, one may express
this orientation using the binary separation vector, $\vect{d}$,
whose magnitude is $d$ and whose direction is that of the binary axis
(to be definite, pointing from the less massive to the more massive
component). With this parameterization,
the projection of the binary orbital motion around its center of mass is
readily facilitated via the time variation of $\vect{d}$. If the time
scale of the event is relatively short compared to the orbital
period of the binary,
then rectilinear relative lens motion,
$\vect{d} = \vect{d}_{t_{\rm c}} +
\dot{\vect{d}}_{t_{\rm c}} (t - t_{\rm c})$, will be an adequate
representation of the actual variation for most applications
\citep[see e.g.][]{PL_MB9741}.
Then, the light curve of a rotating binary event with parallax is
completely specified by eleven independent parameters: 
$\vect{d}_{t_{\rm c}}$, $\dot{\vect{d}}_{t_{\rm c}}$,
$\vect{u}_{t_{\rm c}}$, $\dot{\vect{u}}_{t_{\rm c}}$,
$q$, $\rho_\ast$, and $\pi_{\rm E}$.
However,
one generally chooses to make $t_{\rm c}$ an independent parameter,
such as the time when $\vect{u}_{t_{\rm c}}\cdot\dot{\vect{u}}_{t_{\rm c}}=0$.
In that case, the eleven parameters become
$\vect{d}_{t_{\rm c}}$, $\dot{\vect{d}}_{t_{\rm c}}$,
$u_{t_{\rm c}}$, $\dot{\vect{u}}_{t_{\rm c}}$,
$q$, $\rho_\ast$, $\pi_{\rm E}$, and $t_{\rm c}$.

\begin{figure}
\plotone{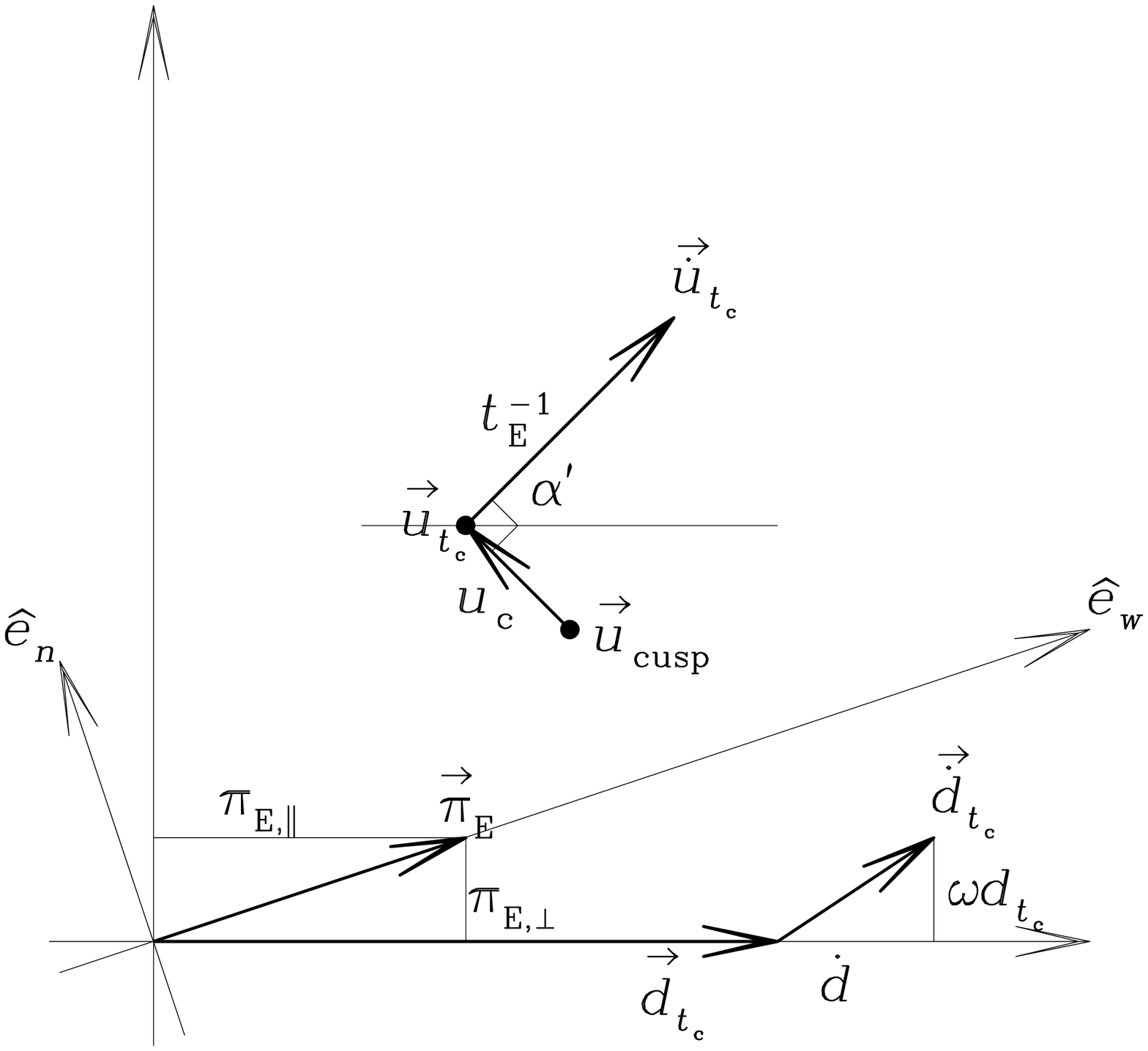}
\caption{\label{fig:geopar}
Geometry used for deriving the transformation shown in Table~\ref{tab:parm}.
The direction of $\vect{d}_{t_{\rm c}}$ is chosen to be the $x$-axis while
$\vect{\pi}_{\rm E}$ lies parallel to the direction of decreasing
ecliptic longitude; $\hat{\vect{e}}_w$. 
The reference time, $t_{\rm c}$ is the time of
the closet approach to the cusp, $\vect{u}_{\rm cusp}$.
}\end{figure}

Although the parameterization described so far is physically motivated,
and mathematically both complete and straightforward,
in practice it is somewhat
cumbersome to implement into the actual fit.
Therefore, we re-formulate the above parameterization
for computational purposes.
For the analysis of \eb, we first choose the reference time
$t_{\rm c}$ as the time of the closest approach of the source to the cusp,
and rotate the coordinate system so that the whole geometry is expressed
relative to the direction of $\vect{d}_{t_{\rm c}}$, 
i.e.\ the binary axis at time $t_{\rm c}$
(see Fig.~\ref{fig:geopar}). We define
the impact parameter for the cusp approach,
$u_{\rm c}$ ($\equiv |\vect{u}_{t_{\rm c}}-\vect{u}_{\rm cusp}|$),
and set
$u_{\rm c}>0$ when the cusp is on the right-hand side of the moving source.
Then,
$\dot{\vect{u}}_{t_{\rm c}}$ is
specified by
$t_{\rm E}^\prime$ ($\equiv|\dot{\vect{u}}_{t_{\rm c}}|^{-1}$),
the instantaneous Einstein timescale at time $t_{\rm c}$,
and by $\alpha^\prime$, the orientation angle of $\dot{\vect{u}}_{t_{\rm c}}$
with respect to $\vect{d}_{t_{\rm c}}$. In addition, we express
$\dot{\vect{d}}_{t_{\rm c}}$ in a polar-coordinate form and 
use the approximation
that both the radial component, $\dot d$, and 
tangential component, $\omega$, are constant.
Under this parameterization, $\dot d$ corresponds to the rate of
expansion ($\dot d > 0$) or contraction ($\dot d < 0$)
of the projected binary separation
while $\omega$ is the angular velocity of the projected binary-axis rotation
on the plane of the sky. Finally, we define the microlens parallax
\emph{vector}, $\vect{\pi}_{\rm E}$, whose magnitude is $\pi_{\rm E}$ and
whose direction is toward ecliptic west (decreasing ecliptic longitude).
In the actual fit, $\pi_{{\rm E},\parallel}$ and $\pi_{{\rm E},\perp}$,
the two projections of $\vect{\pi}_{\rm E}$
along and normal to $\vect{d}_{t_{\rm c}}$, are used as independent
parameters. Table~\ref{tab:parm} summarizes the transformation from
the set of fit parameters ($d_{t_{\rm c}}, q, \alpha^\prime, u_{\rm c},
t_{\rm E}^\prime, t_{\rm c}, \rho_\ast, \pi_{{\rm E},\parallel}, 
\pi_{{\rm E},\perp}, \dot d, \omega$) to the set of the physical parameters
($\vect{d}_{t_{\rm c}}, \dot{\vect{d}}_{t_{\rm c}},
u_{t_{\rm c}}, \dot{\vect{u}}_{t_{\rm c}},
q, \rho_\ast, \pi_{\rm E}, t_{\rm c}$).

\subsection{Terrestrial Baseline Parallax}

In general, the Earth's spin adds a tiny daily wobble of order
$\sim \mbox{R}_\earth/\tilde{r}_{\rm E}$ (eq.~[\ref{eqn:tpar}]),
where R$_\earth$ is the Earth's radius, to the source's relative position
seen from the center of the Earth as expressed in equation~(\ref{eqn:udef}).
Since $\mbox{R}_\earth=4.26\times 10^{-5}\ \mbox{AU}$, this effect is
negligible except when the spatial gradient over the magnification
map is very large, e.g.\ caustic crossings or extreme cusp approaches.
Even for those cases, only the instantaneous offsets are usually
what matters because the source crosses over the region of extreme
gradient with a time scale typically smaller than a day. Hence,
unless the coverage of the crossing from two widely separated observers
significantly overlaps, the effect has been in general ignored
when one models microlensing light curves.

\begin{figure}
\plotone{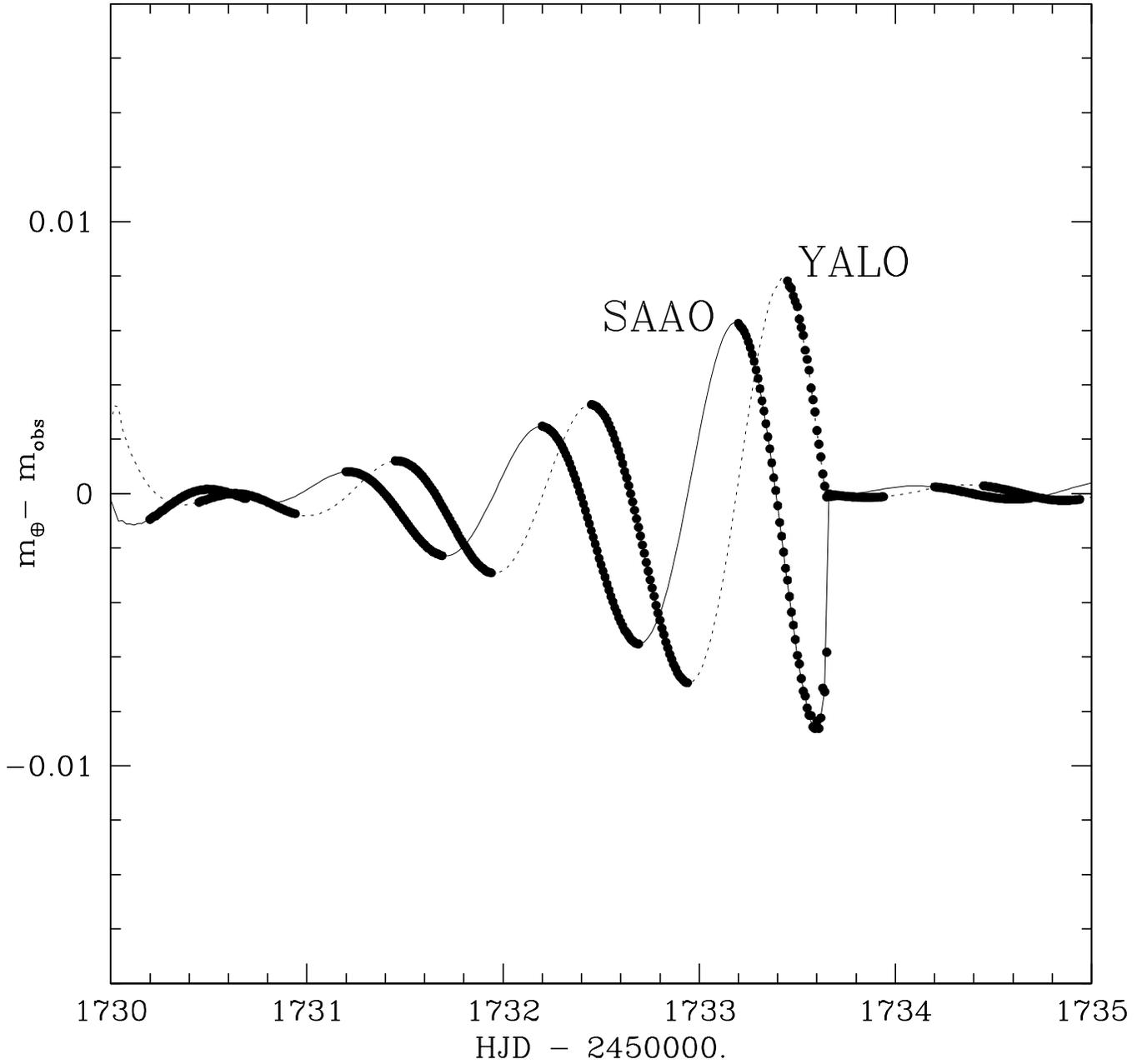}
\caption{\label{fig:tgeo}
Prediction of deviations of light curves for SAAO and YALO from
the geocentric light curve for a chosen model.
The solid curve is the magnitude difference between the
SAAO light curve and the geocentric one, and the dotted curve
is the same for YALO. Nominal night portions (between 6 pm and 6 am local time)
of the light curve are highlighted by overlaid dots.
}\end{figure}

However, in case of \eb, the second caustic crossing lasted four days,
and therefore \emph{daily modulations} of magnifications due to the Earth's
rotation, offset according to the geographic position of each observatory,
may become important, depending on the actual magnitude of the effect
(See also \citealt{Ho99} for a similar discussion on the short time
scale magnification modulation observed from an Earth-orbiting satellite).
\citet{HW95} and \citet{GA99} investigated effects of the terrestrial
baseline parallax, for fold-caustic crossing microlensing events
mainly focused on the instantaneous offsets due to the separation between
observers. They argued that the timing difference
of the trailing limb crossing for observations made from two
different continents could be of the order of tens of seconds
to a minute \citep{HW95} and the magnifications
near the end of exit-type caustic crossings
could differ by as much as a few percent \citep{GA99}.
Suppose that $\phi_2$ is the angle at which the source crosses the caustic,
$A_{\max}$ is the magnification at the peak of the crossing, and
$A_{\rm cc}$ is the magnification right after the end of the crossing.
Then, for the second caustic crossing of \eb, since
$t^\prime_{\rm E} \csc\phi_2 (\mbox{R}_\earth/\tilde r_{\rm E})
\simeq 9\,(0.25/\sin\phi_2)\ \mbox{min}$ and
$\rho_\ast^{-1} (A_{\max}/A_{\rm cc}) (\mbox{R}_\earth/\tilde r_{\rm E})
\simeq 2\times 10^{-2}$,
the time for the end of the second caustic crossing
may differ from one observatory to another by as much as ten minutes
and the magnification difference between them at the end of the crossing
can be larger than one percent, depending on the relative orientation
of observatories with respect to the event at the time of the observations.
Based on a model of the event, we calculate the effect and find that
it causes the magnification modulation of an amplitude as large as
one percent (Fig.~\ref{fig:tgeo}).
In particular, night
portions of the observatory-specific light curves exhibit steeper falls
of flux than would be the case if the event were observed from the Earth's
center. That is, the source appears to move faster during the night
because the reflex of the Earth's rotation is added to the source motion.
This would induce a systematic bias in parameter measurements
if it were not taken into account in the modeling.
We thus include the (daily) terrestrial baseline parallax 
in our model to reproduce
the observed light curve of \eb. Here, we emphasize that this
inclusion requires no new free parameter for the fit once
geographic coordinates of the observatory is specified and R$_\earth$
in units of AU is assumed to be known (see Appendix~\ref{asec:tpar}).

\begin{figure}
\plotone{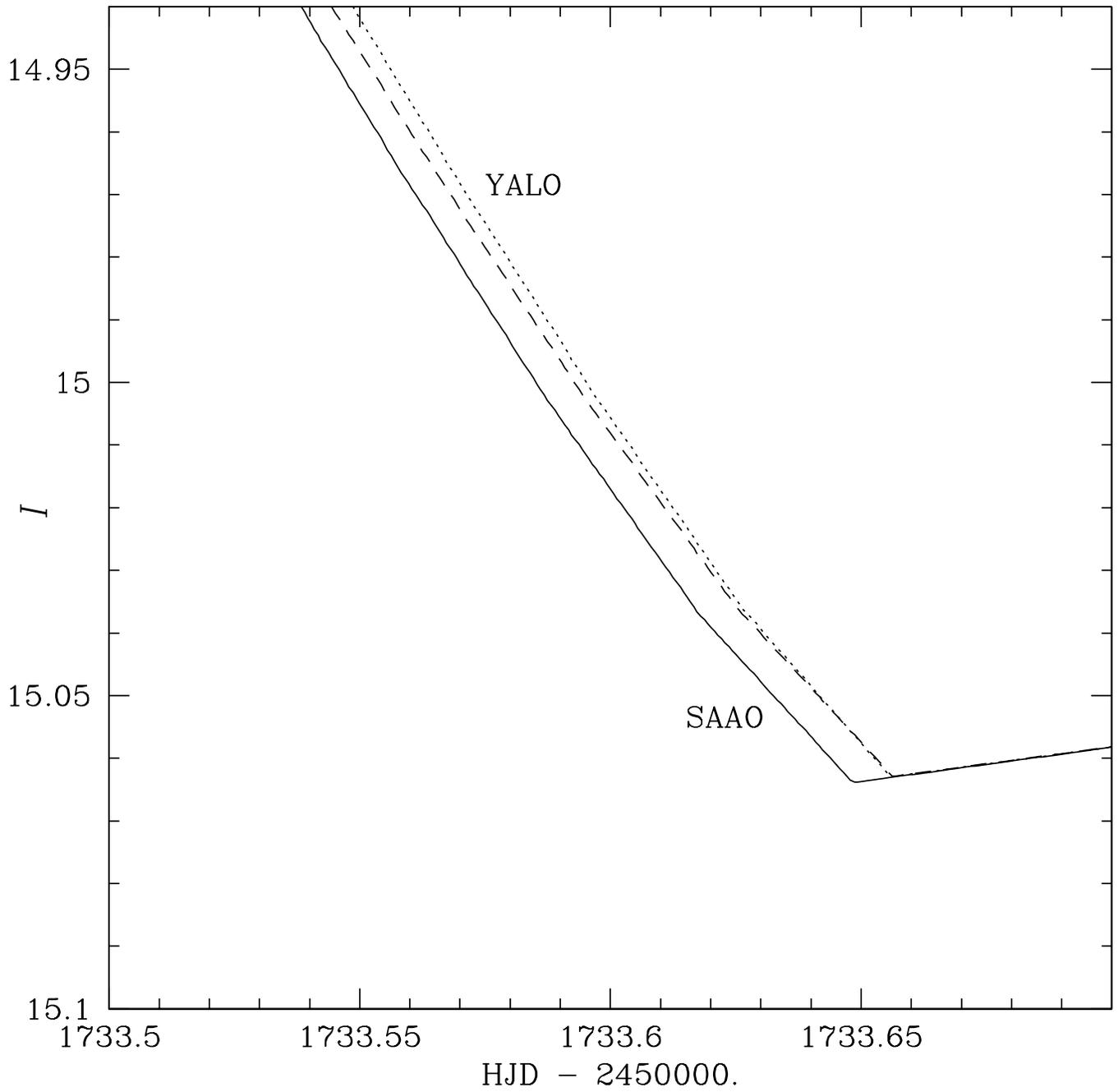}
\caption{\label{fig:tgeo2}
Close-up of model light curves for the end of the second caustic crossing.
The solid curve is modeled for SAAO observations, the dotted curve is
for YALO, and the dashed curve is the geocentric light curve.
The timing of the end of the second crossing for SAAO is earlier than
for YALO by 11 minutes. For comparison purposes, all the light curve are
calculated assuming no blend.
}\end{figure}

In addition, we note the possibility of a simple test of the
terrestrial baseline parallax.
Figure~\ref{fig:tgeo2} shows that the end of the crossing observed
from SAAO is supposed to be earlier than in the geocentric model
-- and earlier than seen from South American observatories.
Unfortunately, near the end of the second caustic crossing
(HJD $\sim 2451733.66$),
PLANET data were obtained only from SAAO near the very end of the night
-- the actual trailing limb crossing is likely to have occurred
right after the end of the night at SAAO,
while YALO was clouded out due to the bad weather at CTIO.\
(The event was inaccessible from telescopes on
Australian sites at the time of trailing limb crossing.)
However, it is still possible to compare the exact timing
for the end of the second crossing derived by other observations
from South American sites with our model prediction and/or
the observation from SAAO. In particular, the EROS collaboration
has published a subset of their observations from
the Marly 1~m at the European Southern Observatory (ESO)
at La Silla, Chile, for the second crossing of \eb\ \citep{EROS_SPEC}.
Comparison between their data and our model/observations may serve
as a confirmation of terrestrial parallax effects.

\subsection{Limb-darkening Coefficients}
Because of the
unprecedentedly long time scale of the second caustic crossing and the
extremely close approach to the cusp,
as well as the high quality of the data,
we adopt a two-parameter limb-darkening law of the form,
\begin{equation}
\label{eqn:limblaw}
S_\lambda (\vartheta)
=
\bar{S}_\lambda
\left[\left(1-\Gamma_\lambda-\Lambda_\lambda\right)
+\frac{3\Gamma_\lambda}{2}\cos\vartheta+
\frac{5\Lambda_\lambda}{4}\cos^{1/2}\vartheta\right]
\,,
\end{equation}
to model the surface brightness profile of the source
Here 
$\bar{S}_\lambda\equiv F_{{\rm s},\lambda}/(\pi \theta_\ast^2)$ is the
mean surface brightness of the source and
$\vartheta$ is the angle between the normal to the stellar surface
and the line of sight, i.e., $\sin\vartheta = \theta/\theta_\ast$ where
$\theta$ is the angular distance to the center of the source.
This is an alternative form of the widely-used
square-root limb-darkening law,
\begin{equation}
\label{eqn:slimb}
S_\lambda (\vartheta)
=
S_\lambda (0)
\left[1-c_\lambda(1-\cos\vartheta)
-d_\lambda(1-\sqrt{\cos\vartheta})\right]
\ .
\end{equation}
However, instead of being normalized to have the same central intensity
$S_\lambda (0)$ as a uniform source, the form we adopt
(eq.~[\ref{eqn:limblaw}]) is normalized to have the same total flux
$F_{{\rm s},\lambda}
=(2\pi\theta_\ast^2)\int_0^1S_\lambda(\vartheta)\sin\vartheta\,d(\sin\vartheta)
$.
That is,
there is no net flux associated with the limb-darkening coefficients. 
The transformation of the coefficients in equation~(\ref{eqn:limblaw})
to the usual coefficients used in equation~(\ref{eqn:slimb}) is given by
\begin{equation}
c_\lambda=\frac{6\Gamma_\lambda}{4+2\Gamma_\lambda+\Lambda_\lambda}
\,;\ \ \
d_\lambda=\frac{5\Lambda_\lambda}{4+2\Gamma_\lambda+\Lambda_\lambda}
\ .\end{equation}
Note that, although we fit limb-darkening,
the detailed discussion of its measurement and error analysis
will be given elsewhere.

\section{Measurement of 
the Projected Einstein Radius
\label{sec:pie}}

\begin{figure}
\plotone{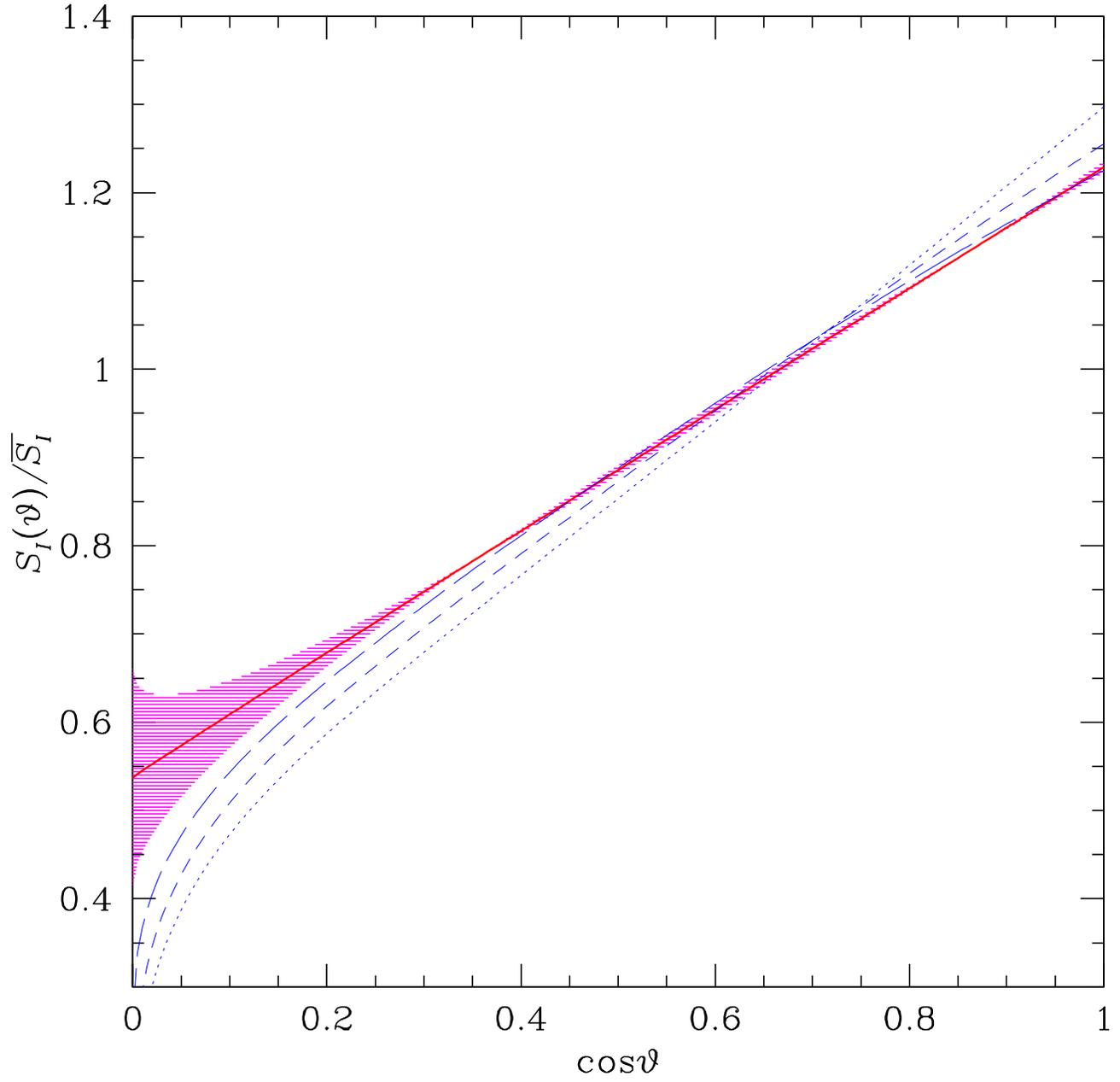}
\caption{\label{fig:prof}
Surface brightness profile of the source star.
The thick solid curve is the prediction indicated by the best
fit model (see Table~\ref{tab:limb}). In addition, the variation of
profiles with the parameters allowed to deviate
by 2-$\sigma$ along the direction of the principal conjugate
is indicated by a shaded region. For comparison, also shown are
theoretical profiles taken from \cite{Cl00}.
The stellar atmospheric model parameters for them are
$\log g=1.0$, [Fe/H]$=-0.3$, and 
$T_{\rm eff}=3500\ \mbox{K}$ (dotted curve),
$4000\ \mbox{K}$ (short dashed curve), 
$4500\ \mbox{K}$ (long dashed curve).
Note that the effective temperature of the source is reported to be
$4500\pm 250\ \mbox{K}$ by \citet{VLT} and $3800\pm 200\ \mbox{K}$ 
by \citet{UVES}.
}\end{figure}

Table~\ref{tab:fmod} gives the parameters describing the best-fit microlens
model (see Appendix~\ref{asec:fit} for details of modeling and
Appendix~\ref{asec:covar} for the discussion on the error determination)
for the PLANET $I$-band observations of \eb.
We also transform the fit parameters
to the set of parameters introduced in \S~\ref{ssec:par}.
In Table~\ref{tab:limb}, we provide the result of limb-darkening
coefficient measurements. The measurements of two coefficients, $\Gamma_I$
and $\Lambda_I$ are highly anti-correlated so that the uncertainty along
the major axis of error ellipse is more than 50 times larger than
that along its minor axis. While this implies that the constraint
on the surface brightness profile
derived by the caustic crossing light curve
is essentially one dimensional, its natural basis is neither
the linear nor the square-root parameterized form.
We plot, in Figure~\ref{fig:prof},
the surface brightness profile indicated by the fit and compare these
with theoretical calculations taken from \citet{Cl00}.
The figure shows that allowing the profile parameters to vary by
2-$\sigma$ does not have much effect on the central slope, but
cause a large change in the behavior
near the limb. We speculate that this may be related to the
specific form of the time sampling over the stellar disk, but
more detailed analysis regarding
limb darkening is beyond the scope of the present paper and
will be discussed elsewhere.

\begin{figure}
\plotone{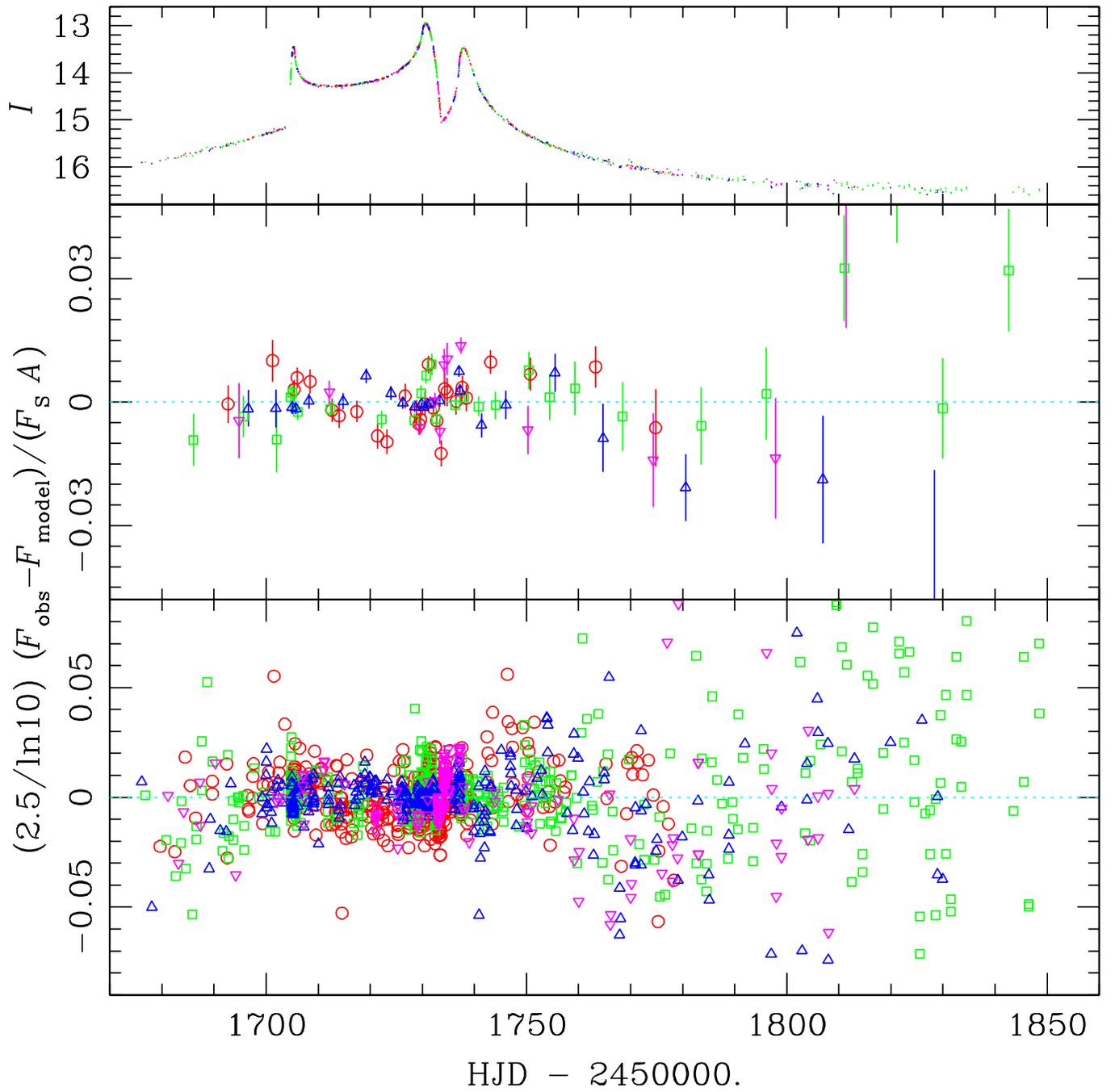}
\caption{\label{fig:resid}
``Magnitude'' residuals from PLANET model of \eb.
Symbols are the same as in Fig.~\ref{fig:ltc}
The top panel shows the light curve corresponding to the time of
observations, the middle panel shows the averaged residuals
from 15-sequential observations, and the bottom panel shows the
scatters of individual residual points.
}\end{figure}

\begin{figure}
\plotone{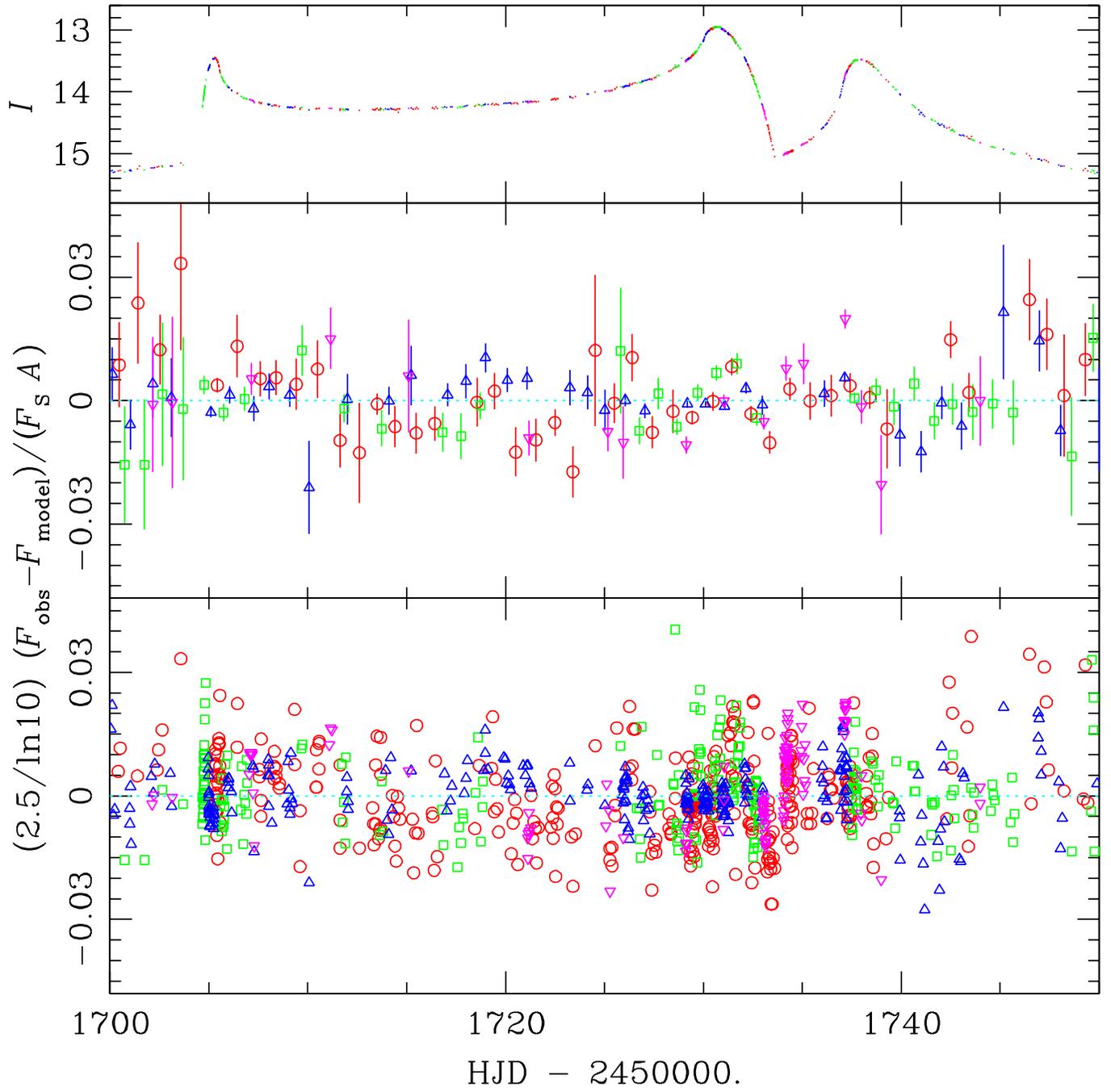}
\caption{\label{fig:res2}
Same as Fig.~\ref{fig:resid} but focuses mainly on the
``anomalous'' part of the light curve. The middle panel
now shows the daily averages of residuals.
}\end{figure}

\begin{figure}
\plotone{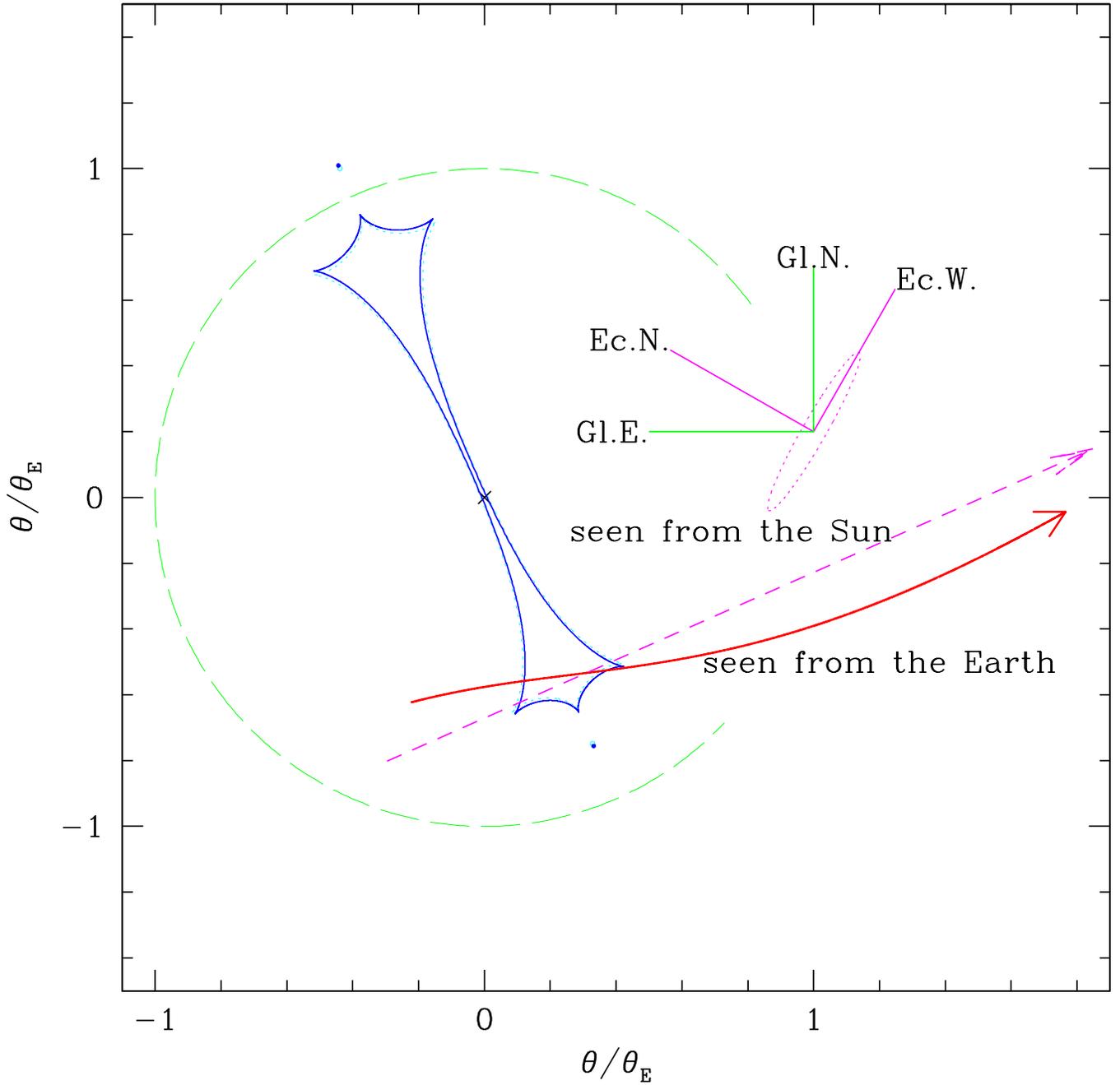}
\caption{\label{fig:geo}
Geometry of the event projected on the sky. Left is Galactic east,
up is Galactic north.
The origin is the center of mass of the binary lens. The trajectory
of the source relative to the lens is shown as a thick solid curve
while the short-dashed line shows the relative proper
motion of the source seen from the Sun. The lengths of these trajectories
correspond to the movement over six months 
between HJD $= 2451670$ and HJD $= 2451850$.
The circle drawn with long dashes indicates the Einstein ring, and
the curves within the circle are the caustics at two different times. The
solid curve is at $t=t_{\rm c}$ while the dotted curve
is at the time of the first crossing. The corresponding locations of the two
lens components are indicated by filled ($t=t_{\rm c}$) and open (the
first crossing) dots. The lower dots represent the more 
massive component of the binary.
The ecliptic coordinate basis is also overlayed with the elliptical
trajectory of $\pi_{\rm E} \vect{\varsigma}$ over the year.
}\end{figure}

\begin{figure}
\plotone{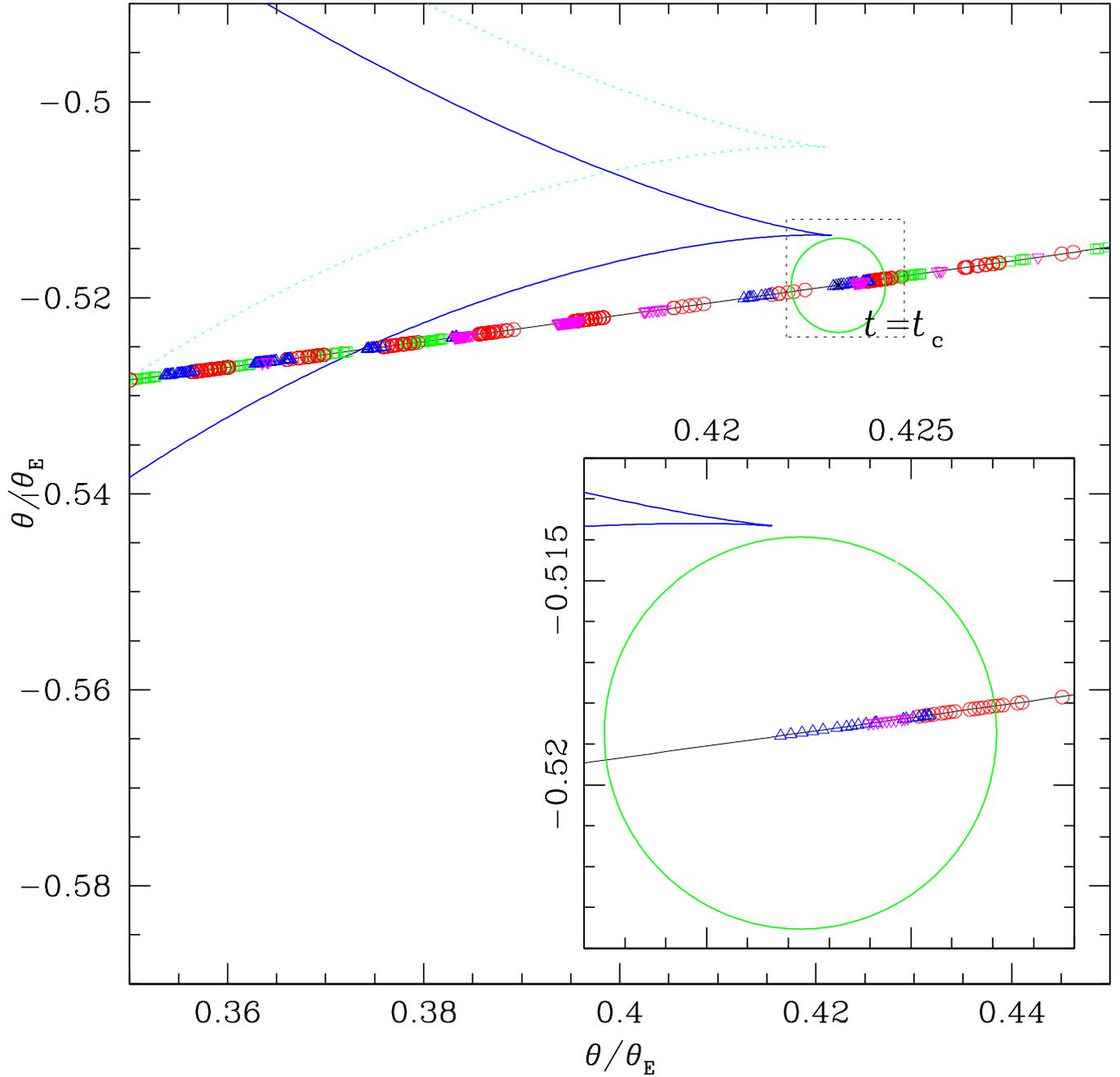}
\caption{\label{fig:zoom}
Close-up of Fig.~\ref{fig:geo} around the cusp approach. The source
at the closest approach ($t=t_{\rm c}$) is shown as a circle.
The solid curve is the caustic at $t=t_{\rm c}$ while the dotted curve
is the caustic at the time of the first crossing.
The positions of the source center at the time of each of the observations
are also shown by symbols
(same as in Figs.~\ref{fig:ltc}, \ref{fig:resid}, \ref{fig:res2})
that indicate the observatory.
For the close-up panel, only those points
that were excluded from the fit because of numerical problems in the
magnification calculation (see \S\S~\ref{sec:dat} and \ref{sec:pie})
are shown. Note that the residuals for all points (included these
excluded ones) are shown in Figs.~\ref{fig:resid} and \ref{fig:res2}.
}\end{figure}

Figures~\ref{fig:resid} and \ref{fig:res2} show ``magnitude residuals'',
$(2.5/\ln 10) [\Delta F/(A F_{\rm S})]$,
from our best model.
Note in particular that the residuals for the points near
the cusp approach (HJD $= 2451736.944$) 
that were excluded from the fit do not differ qualitatively
from other residuals. It is true that the mean residual for Perth 
on this night (beginning $4.0\ \mbox{hrs}$ after $t_{\rm c}$
and lasting $2.3\ \mbox{hrs}$) was about 2\% high.
However, the Canopus data, which span the whole
cusp approach from $1.2\ \mbox{hrs}$ before 
until $7.7\ \mbox{hrs}$ after $t_{\rm c}$,
agree with the model to within 0.5\%.
Moreover, the neighboring SAAO and YALO points also show
excellent agreement. See also Fig.~\ref{fig:geo} and especially
Fig.~\ref{fig:zoom}.

From the measured microlens parallax ($\pi_{\rm E} = \PIE$), the projected
Einstein radius is (eq.~[\ref{eqn:pie}]);
\begin{equation}
\tilde r_{\rm E} =
\frac{1\ \mbox{AU}}{\pi_{\rm E}} =
\RETILDE
\ .\end{equation}
We also derive $t_{\rm E}$,
the heliocentric 
Einstein timescale;
\begin{mathletters}
\begin{equation}
\vect{\mu}_{\rm E}= 
\dot{\vect{u}}_{t_{\rm c}} + \pi_{\rm E} \dot{\vect{\varsigma}}_{t_{\rm c}}
\,,\end{equation}
\begin{equation}
t_{\rm E} \equiv |\vect{\mu}_{\rm E}|^{-1}
= \TE
\,,\end{equation}
\end{mathletters}
and the direction of $\vect{\mu}_{\rm E}$ is $\MUEANEC$ from ecliptic
west to south. Since, toward the direction of the event, the Galactic disk
runs along 60\fdg2 from ecliptic west to south, $\vect{\mu}_{\rm E}$
points to $\MUEANGL$ north of the Galactic plane.
The overall geometry of the model is shown in Figure~\ref{fig:geo},
and Figure~\ref{fig:zoom} shows a close-up of the geometry
in the vicinity of the cusp approach ($t\sim t_{\rm c}$).
Next, the projected velocity $\tilde{\vect{v}}$ on the observer plane is
\begin{equation}
\label{eqn:vtilval}
\tilde{\vect{v}} = \tilde r_{\rm E} \vect{\mu}_{\rm E} =
\left(
\VTILDEE
,
\VTILDEN
\right)\ \mbox{km s$^{-1}$}
\ .\end{equation}
Here the vector is expressed in Galactic coordinates.
We note that the positive $x$-direction is the direction of the
Galactic rotation -- i.e.\ the apparent motion of the Galactic
center seen from the Local Standard of Rest (LSR) is in the negative
$x$ direction -- while the
positive $y$-direction is toward the north Galactic pole so that
the coordinate basis is left-handed.

\section{Measurement of
the Angular Einstein Radius
\label{sec:thetae}}

The angular radius of the source, $\theta_\ast$, is determined by placing the
source on an instrumental CMD and finding its
offset relative to the center of the red giant clump,
whose de-reddened and calibrated position toward the Galactic bulge
are known from the literature.
For this purpose, we use the data obtained from YALO.
We find the instrumental $I$ magnitude of the (de-blended) source
$I_{\rm s}$ from the fit to the light curve.
To determine the color, we first note that,
except when the source is near (and so partially resolved by) a caustic,
microlensing is achromatic. That is, the source is equally magnified
in $V$ and $I$: $F_V = F_{V,{\rm b}} + A F_{V,{\rm s}}$
and $F_I = F_{I,{\rm b}} + A F_{I,{\rm s}}$.
Hence, we assemble pairs of $V$ and $I$ points observed within 30 minutes
(and excluding all caustic-crossing and cusp-approach data) and fit these
to the form $F_V = a_1 + a_2 F_I$.  We then find
$(V-I)_{\rm s, inst} = -2.5\log a_2$.  We also find the magnitude of the blend
$I_{\rm b}$ from the overall fit, and solve for the color of the blend
$(V-I)_{\rm b, inst} = -2.5\log [a_2 + (a_1/F_{I,{\rm b}})]$.

We then locate the source on the instrumental CMD and measure its
offset from the center of the red giant clump;
$\Delta (V-I) = 0.276\pm 0.010$ and $\Delta I = 0.33\pm 0.02$.
Here, the quoted uncertainty includes terms for both the clump center and
the source position. Finally, using the calibrated and de-reddened
position of the red clump, 
$[(V-I)_0,I_0]=[1.114\pm 0.003, 14.37\pm 0.02]$ \citep{OGLE_CLUMP},
the source is at $(V-I)_0 = 1.390\pm 0.010$, $I_0=14.70\pm 0.03$.
The procedure does not assume any specific reddening law for determining
the de-reddened source color and magnitude, but may be subject to a
systematic error due to differential reddening across the field.

\begin{figure}
\plotone{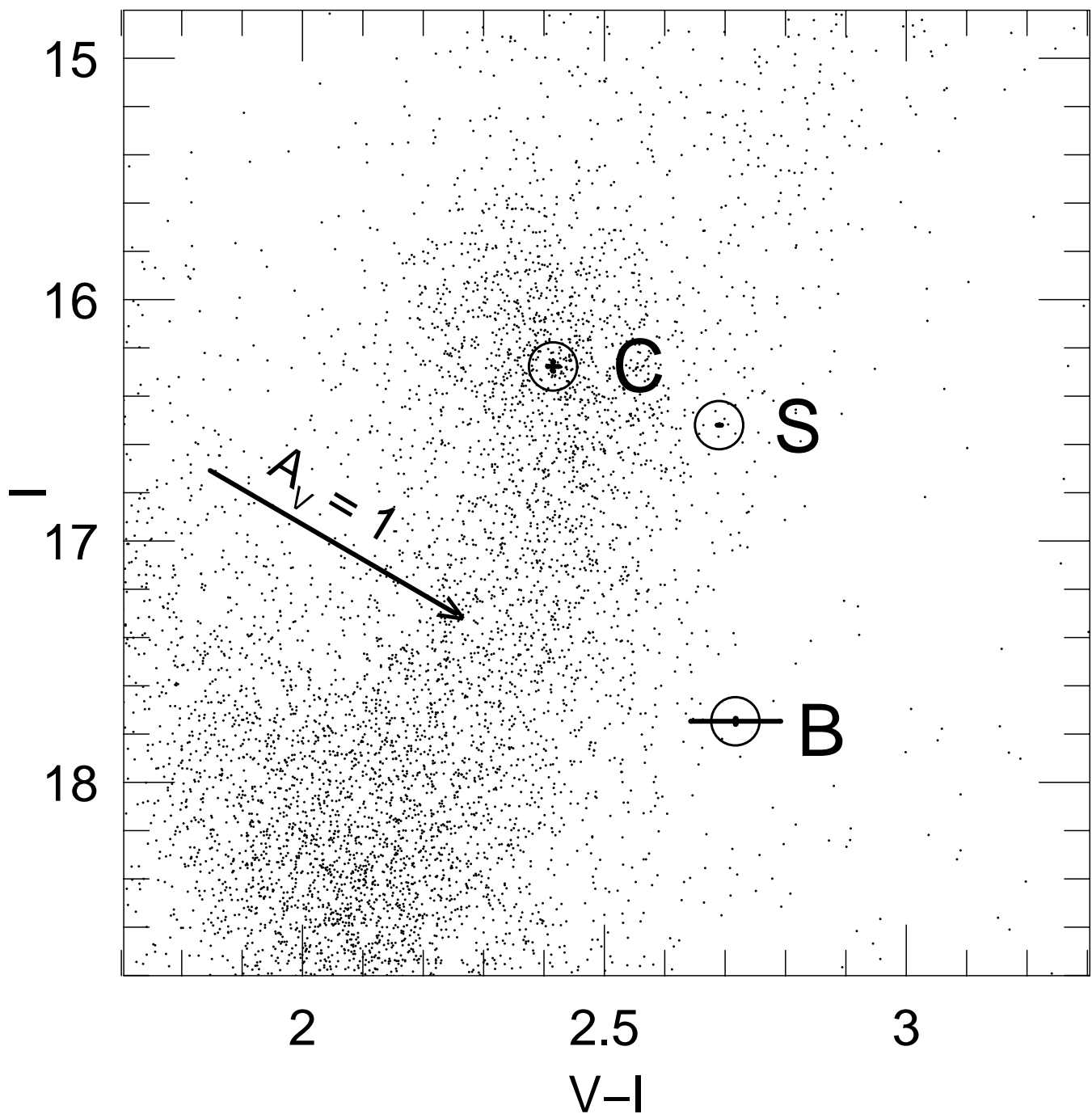}
\caption{\label{fig:cmd}
Calibrated CMD for the field around \eb, observed from YALO.
The arrow shows the reddening vector.
The positions of the source (S), blend (B), and the center of
the red giant clump (C) are denoted by capital letters.
}\end{figure}

We also perform a photometric calibration for observations made
at SAAO. The calibration was derived from images obtained on three
different nights by observing \citet{La92} standards in the 2000 season,
and the resulting transformation
equation were confirmed with a further night's observations in 2001.
Applying the photometric transformation on the fit result for the
source flux, we obtain the -- unmagnified -- source magnitude of
$I=16.602$ in the standard (Johnson/Cousins) system, while the
standard source color is measured to be $V-I=2.69$.
By adopting the photometric offset between SAAO and YALO
instrumental system determined by the fit, we are able to derive
a ``calibrated'' YALO CMD for the field around \eb\ (Fig.~\ref{fig:cmd}). 
The positions of the source (S), blend (B),
and clump center (C) are also overlaid on the CMD.
The CMD also implies that the reddening toward the field is
$A_I=1.90$ and $E(V-I)=1.30$, which yields a reddening law,
$R_{VI}=A_V/E(V-I)=2.46$, nominally
consistent with the generally accepted $R_{VI}=2.49\pm 0.02$ \citep{St96}.
If we adopt a power-law extinction model
$A_\lambda\propto\lambda^{-p}$, then $R_{VI}=2.46$ corresponds to
$p=1.37$. For this index, $E(\bv)/E(V-I)=0.838$,
which predicts a lower extinction than
the spectroscopically determined reddening of $E(\bv)=1.30\pm 0.05$
by \citet{VLT}.

Both the source and the blend lie redward of the main population
of stars in the CMD. One possible explanation is that the line of
sight to the source is more heavily reddened than average
due to differential reddening across the field.
Inspection of the images does indicate significant differential reddening,
although from the CMD itself it is clear that only a small minority of
stars in the field could suffer the additional $\Delta A_V \sim 0.7$
that would be necessary to bring the source to the center of the clump.
There is yet another indication that the source has average extinction
for the field -- i.e.\ the same or similar extinction as the clump center.
The de-reddened (intrinsic) color derived on this assumption, $(V-I)_0 = 1.39$
is typical of a K3III star \citep{BB88}, which is in good
agreement with the spectral type determined by \citet{VLT}.
On the other hand, if the source were a reddened clump giant with
$(V-I)_0\sim 1.11$, then it should be a K1 or K2 star.

We then apply the procedure of \citet{PL_MB9741} to derive the angular
source radius $\theta_\ast$ from its de-reddened color and magnitude:
first we use the color-color relations of \citet{BB88} to convert from
$(V-I)_0$ to $(V-K)_0$, and then we use the empirical relation
between color and surface brightness to obtain $\theta_\ast$
\citep{vB99}.
From this, we find that
\begin{eqnarray}
\label{eqn:thstar}
\theta_\ast 
&=&
6.62\pm 0.58\ \mbox{$\mu$as}
\nonumber\\&=&
1.42\pm 0.12\ \mbox{R$_\sun$ kpc$^{-1}$}
\,,
\end{eqnarray}
where the error is dominated by
the 8.7\% intrinsic scatter in the relation of \citet{vB99}.
Alternatively, we use the different calibration derived for
K giants by \citet*{BBH99} and obtain $\theta_\ast=6.47\ \mbox{$\mu$as}$,
which is consistent with equation~(\ref{eqn:thstar}). Finally,
we note that if the source were actually a clump giant that suffered
$\Delta A_V=0.7$ extra extinction,
its angular size would be $7.08\ \mbox{$\mu$as}$.
Since we consider this scenario unlikely, and since in any event the
shift is smaller than the statistical error, we ignore this possibility.

From this determination of $\theta_\ast$ and the value,
$\rho_\ast=(\RHOSTAR)\times 10^{-3}$, determined 
from the fit to the light curve,
we finally obtain,
\begin{equation}
\theta_{\rm E} = \frac{\theta_\ast}{\rho_\ast} = \THETAE
\,,\end{equation}
where the error is again dominated by the intrinsic scatter in the
relation of \citet{vB99}.

\section{The Lens Mass and the Lens-Source Relative Proper Motion}

By combining the results obtained in \S\S~\ref{sec:pie} and \ref{sec:thetae},
one can derive several key physical parameters, including
the lens mass and the lens-source relative 
parallax and proper motion;
\begin{equation}
M = \frac{c^2}{4G} \tilde r_{\rm E} \theta_{\rm E}
= \MASS
\,,\end{equation}
\begin{mathletters}
\begin{equation}
D_{\rm rel} = \frac{\tilde r_{\rm E}}{\theta_{\rm E}}
= \DREL
\,;\end{equation}
\begin{equation}
\pi_{\rm L} - \pi_{\rm S} = \frac{\mbox{1 AU}}{D_{\rm rel}} 
= \PIREL
\ .\end{equation}
\end{mathletters}
\begin{eqnarray}
|\vect{\mu}_{\rm S}-\vect{\mu}_{\rm L}| = \theta_{\rm E} |\vect{\mu}_{\rm E}|
&=&
\MURELA
\nonumber\\&=&
\MURELB
\ .\end{eqnarray}

For the binary mass ratio of the best-fit model, $q=0.749$,
the masses of the individual components are $0.350\ \mbox{M$_\sun$}$
and $0.262\ \mbox{M$_\sun$}$, both of which are consistent with
the mass of typical mid-M dwarfs in the Galactic disk.
Using the mass-luminosity relation of \citet{HM93}, and adopting
$M_V=2.89+3.37(V-I)$, the binary then has a combined color and
magnitude $M_I=8.2\pm 0.2$ and $(V-I)=2.54\pm 0.08$.
Since $D_{\rm rel}^{-1} = D_{\rm L}^{-1} - D_{\rm S}^{-1}$,
$D_{\rm rel}$ is an upper limit for $D_{\rm L}$,
i.e.\ the lens is located in the Galactic disk within $2.6\ \mbox{kpc}$
of the Sun.
Furthermore, an argument based on the kinematics (see \S~\ref{ssec:kin})
suggests that $D_{\rm S}\ga 8\ \mbox{kpc}$ so that it is
reasonable to conclude that $D_{\rm L}\sim 2\ \mbox{kpc}$ ($m-M\simeq 11.5$).
Hence, even if the binary lens lay in front of all the extinction
along the line of sight, it would be $\sim 2\ \mbox{mag}$ fainter than
the blend (B), and so cannot contribute
significantly to the blended light.
However, the proximity of the lens to the observer is
responsible for the event's long time scale and quite large
parallax effect.

\section{The Kinematic Constraints on the Source Distance
\label{ssec:kin}}
The projected velocity $\tilde{\vect{v}}$ (eq.~[\ref{eqn:vtilval}]) 
is related to the kinematic parameters of the event by
\begin{mathletters}
\begin{eqnarray}
\label{eqn:vtilde}
\tilde{\vect{v}} 
&=&
D_{\rm rel}\left(\vect{\mu}_{\rm S} - \vect{\mu}_{\rm L}\right)
\nonumber\\&=&
\left(D_{\rm rel}\frac{\vect{v}_{\rm S}}{D_{\rm S}}
-D_{\rm rel}\frac{\vect{v}_{\rm L}}{D_{\rm L}} + \vect{v}_\sun
\right)_\perp
\,,\end{eqnarray}{\rm
where the final subscript, $\perp$, serves to
remind us that all velocities are projected onto the plane of the sky.  
Writing $\vect{v}_\sun=\vect{v}_{\rm rot} + \vect{v}_{\sun,{\rm p}}$ 
and $\vect{v}_{\rm L}=\vect{v}_{\rm rot} + \vect{v}_{\rm L,p}$
as the sum of the Galactic rotation and the peculiar velocities and
eliminating $D_{\rm L}$ in favor of $D_{\rm rel}$ and $D_{\rm S}$,
equation~(\ref{eqn:vtilde}) can be expressed as
}\begin{equation}
\label{eqn:kin}
\tilde{\vect{v}} =
\left[\frac{D_{\rm rel}}{D_{\rm S}}(\vect{v}_{\rm S} - \vect{v}_{\rm rot})
- \left(1+\frac{D_{\rm rel}}{D_{\rm S}}\right)\vect{v}_{\rm L,p}
+ \vect{v}_{\sun,{\rm p}}
\right]_\perp
\ .\end{equation}
\end{mathletters}
For a fixed $D_{\rm S}$ and with a known Galactic kinematic model,
the distribution function of the expected value for $\tilde{\vect{v}}$
can be derived from equation~(\ref{eqn:kin}). 
The measured value of $\tilde{\vect{v}}$ can
therefore be translated into the relative likelihood, ${\mathcal{L}}$
for a given $D_{\rm S}$ and the assumed kinematic model,
\begin{mathletters}
\begin{equation}
-2\ln{\mathcal{L}} =
(\tilde{\vect{v}}-\langle\tilde{\vect{v}}\rangle)
\cdot\mathbf{C}^{-1}\cdot
(\tilde{\vect{v}}-\langle\tilde{\vect{v}}\rangle)
+\ln|\mathbf{C}|+{\rm constant}
\,,\end{equation}
\begin{equation}
\langle\tilde{\vect{v}}\rangle =
x_{\rm S}(\langle\vect{v}_{{\rm S},\perp}\rangle - \vect{v}_{\rm rot}) -
(1+x_{\rm S})\langle\vect{v}_{{\rm L,p},\perp}\rangle +
\vect{v}_{\sun,{\rm p},\perp}
\,,\end{equation}
\begin{equation}
\mathbf{C}=
x_{\rm S}^2 \mathbf{C}[\vect{v}_{{\rm S},\perp}] + 
(1+x_{\rm S})^2 \mathbf{C}[\vect{v}_{{\rm L},\perp}] +
\mathbf{C}[\tilde{\vect{v}}]
\,,\end{equation}
\end{mathletters}
where $\mathbf{C}[\vect{v}_{{\rm S},\perp}]$ and 
$\mathbf{C}[\vect{v}_{{\rm L},\perp}]$ are the dispersion tensors
for the source and the lens transverse velocity,
$\mathbf{C}[\tilde{\vect{v}}]$ is the covariance tensor for
the measurement of $\tilde{\vect{v}}$,
and $x_{\rm S} = D_{\rm rel}/D_{\rm S}$.

\begin{figure}
\plotone{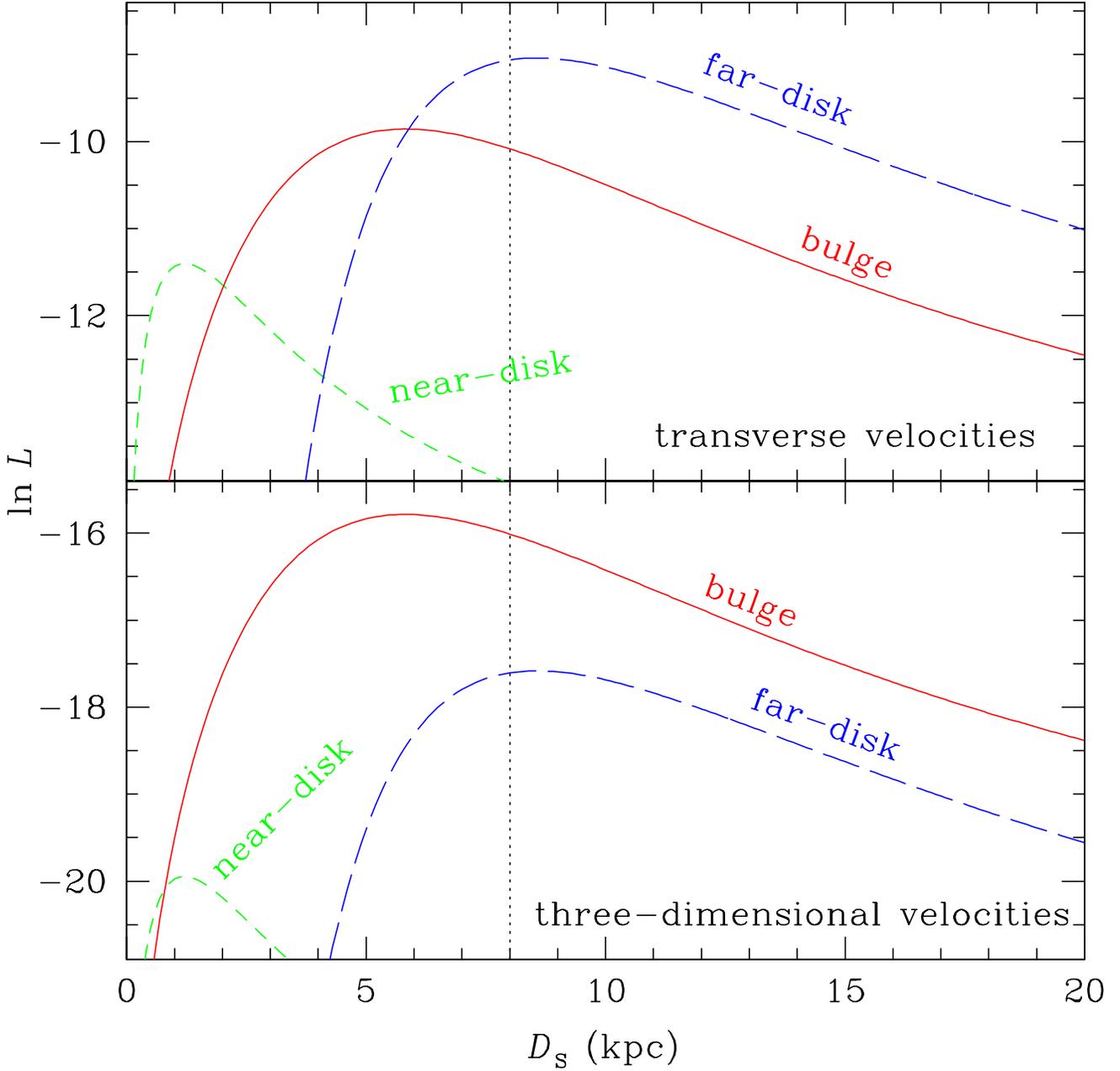}
\caption{\label{fig:likelihood}
Kinematic likelihood for $\tilde{\vect{v}}$ as a function of $D_{\rm S}$.
The three curves are for different distributions of the source velocity:
near-disk-like; $\vect{v}_{\rm S}=\vect{v}_{\rm rot}+\vect{v}_{\rm S,p}$
(short-dashed line), bulge-like; 
$\vect{v}_{\rm S}=\vect{v}_{\rm S,p}$ (solid line),
and far-disk-like; $\vect{v}_{\rm S}=-\vect{v}_{\rm rot}-\vect{v}_{\rm S,p}$
(long-dashed line). The top panel shows the likelihood derived using only
the two-dimensional projected velocity information while in the bottom
panel, the likelihood also includes the radial velocity information derived
from the high resolution spectra of \citet{Keck}.
}\end{figure}

We evaluate ${\mathcal{L}}$ as a function of $D_{\rm S}$ assuming
three different distributions for $\vect{v}_{{\rm S},\perp}$ which
respectively correspond to source being in the near disk
($\langle\vect{v}_{{\rm S},\perp}\rangle=
\vect{v}_{\rm rot}+\langle\vect{v}_{{\rm S,p},\perp}\rangle$),
bulge ($\langle\vect{v}_{{\rm S},\perp}\rangle=
\langle\vect{v}_{{\rm S,p},\perp}\rangle$), and the far disk
($\langle\vect{v}_{{\rm S},\perp}\rangle=
-\vect{v}_{\rm rot}-\langle\vect{v}_{{\rm S,p},\perp}\rangle$).
Adopting the Galactic rotational velocity,
$\vect{v}_{\rm rot} = (220,0)\ \mbox{km s$^{-1}$}$,
the solar motion, 
$\vect{v}_{\sun,{\rm p},\perp} = (5.2,7.2)\ \mbox{km s$^{-1}$}$
\citep{BM98},
and the kinematic characteristics
of the lens and the source given in Table~\ref{tab:kin}, we derive
the relative likelihood as a function of 
source position (Fig.~\ref{fig:likelihood}),
and find that the measured value of $\tilde{\vect{v}}$ 
(eq.~[\ref{eqn:vtilval}]) mildly favors the far disk 
over the bulge as the location of the source
by a factor of $\sim 2.3$. The near disk location is quite
strongly disfavored (by a factor of $\sim 10.6$ compared to
the far disk, and by a factor of $\sim 4.7$ compared to the bulge).

The complete representation of the likelihood for the source location
requires one to consider all the available constraints relevant to the
source distance. In particular, these include the radial (line-of-sight)
velocity measurement and the number density of stars constrained by
the measured color and brightness (or luminosity) along the line of sight.
Although one may naively expect that a bulge location is favored by
the high density of stars in the bulge, which follows from the fact
that the line of sight passes within 400 pc of the Galactic center,
it is not immediately obvious how the disk contribution compares to
the bulge stars for the specific location of the source on the CMD
(Fig.~\ref{fig:cmd}).
Even with no significant differential reddening over the field,
the particular source position,
which is both fainter and redder than the center of the clump,
can be occupied by 
relatively metal-rich (compared to the bulge average) first ascent giants
in either the bulge or the far disk. Because first ascent giants
become redder with increasing luminosity, the source must have
higher metallicity if it lies in the bulge than in the more distant disk.
There exist two estimate of the source metal abundance: [Fe/H]$=-0.3\pm 0.1$
by \citet{VLT} based on VLT FORS1 spectra and [Fe/H]$=-0.3\pm 0.3$ by
\citet{UVES} based on VLT UVES spectra. However, to incorporate these
measurement into a likelihood estimate would require a more through
understanding of all the sources of error as well as a detailed model of
the metallicity distribution of the bulge and the disk.
For this reason, we defer the proper calculation
of the likelihood in the absence of any definitive way to incorporate
the specific density function, and restrict ourselves to the kinematic
likelihood.

We determine the
radial velocity of the source from Keck HIRES spectra of \citet{Keck},
kindly provided to us by R. M. Rich.
We find the line-of-sight velocity
to be $\sim -100\ \mbox{km s$^{-1}$}$ (blueshift; heliocentric)\footnote{
Recently \citet{UVES} reported a radial velocity measurement
of $-191\ \mbox{km s$^{-1}$}$ for the source star of \eb. At this point,
we do not know what is the reason for this discrepancy.}, which
translates to $\sim -90\ \mbox{km s$^{-1}$}$ Galactocentric radial velocity
accounting for the solar motion of $10\ \mbox{km s$^{-1}$}$.
The derived radial velocity strongly favors bulge membership of the
source since it is three times larger than the
line-of-sight velocity dispersion for disk K3 giants, but is consistent
with the motions of typical bulge stars (c.f.\ Table~\ref{tab:kin}).
Because the correlation between the radial and
the transverse velocity for K3 giants either in the disk or in the bulge
is very small, the likelihood for the radial velocity can be
estimated independently from the likelihood for the transverse velocity,
and the final kinematic likelihood is a simple product of two components.
We find that the final kinematic likelihood including the radial
velocity information indicates that the source star belongs to the
bulge (Fig.~\ref{fig:likelihood}): the likelihood for the bulge
membership is about six times larger than that for the far disk membership.

Finally, we note that,
if the source lay in the far disk, it should have
an \emph{additional} retrograde proper motion of 
$\sim 4\ \mbox{mas yr$^{-1}$}\times(10\ \mbox{kpc}/D_{\rm S})$
with respect to the bulge stars,
which should be measurable using adaptive optics or 
the \emph{Hubble Space Telescope}.

\section{Consistency between the Lens Mass and the Binary Orbital Motion}

\begin{figure}
\plotone{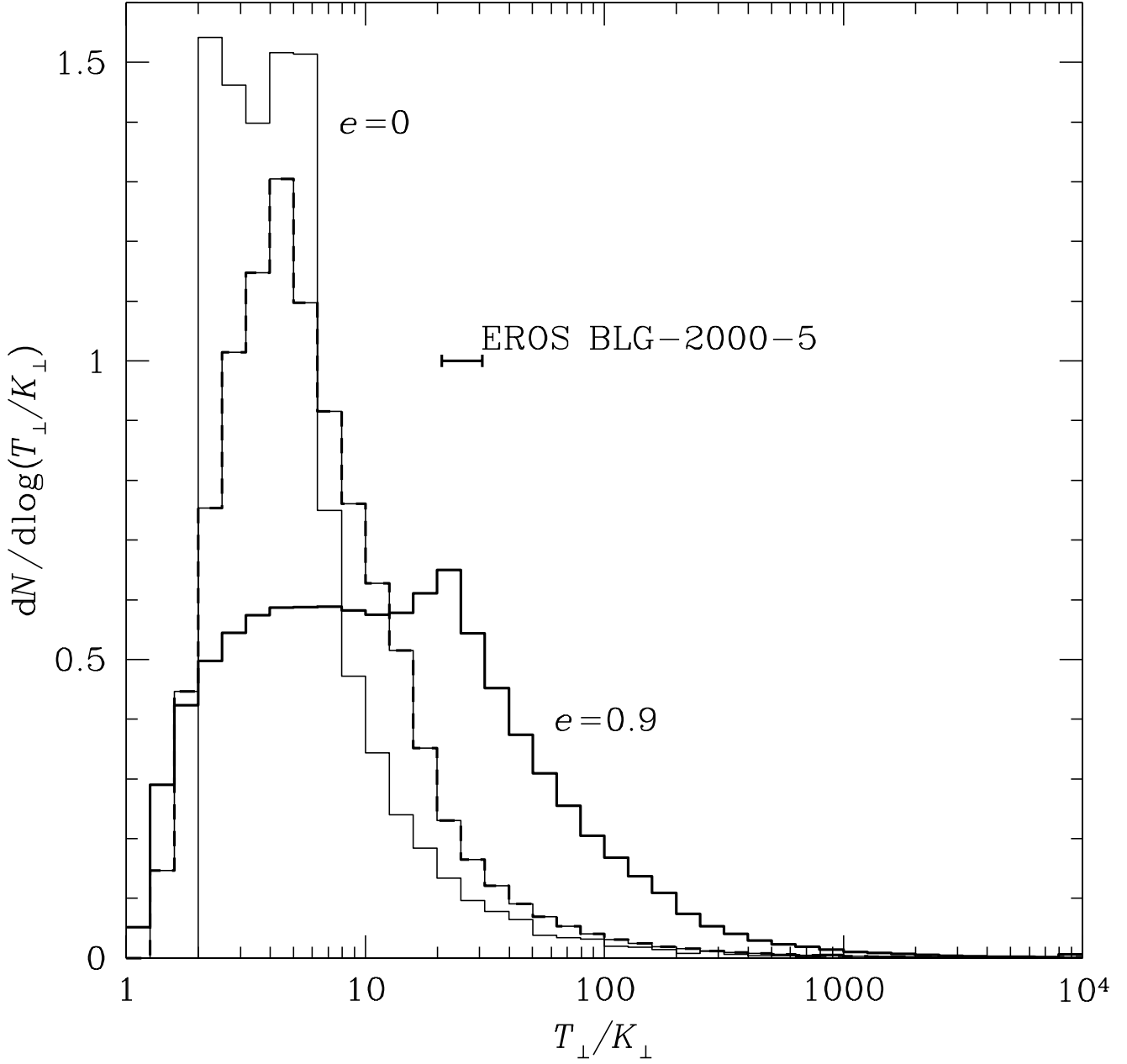}
\caption{\label{fig:rot}
Distributions of the ratio of transverse potential energy,
$|T_\perp|=[q/(1+q)^2]GM^2/r_\perp$ to transverse kinetic energy,
$K_\perp=[q/(1+q)^2]M v_\perp^2/2$, for binaries seen at random times
and random orientations, for three different eccentricities,
$e=0,0.5,0.9$.  Also shown is the 1 $\sigma$ allowed range for \eb.
Non-circular orbits are favored.
}\end{figure}

For the derived parameters, we find a projected binary lens separation
$r_\perp=d\,D_{\rm L}\theta_{\rm E}=5.52\ \mbox{AU}$
$(5.25\ \mbox{AU})$
and transverse orbital speed
$v_\perp=[\dot d^2 + (\omega d)^2]^{1/2}D_{\rm L}\theta_{\rm E}
=2.76\ \mbox{km s$^{-1}$}$ $(2.62\ \mbox{km s$^{-1}$})$
for $D_{\rm S} = 10\ \mbox{kpc}$ $(8\ \mbox{kpc})$.
We can now define the transverse kinetic and potential energies
$K_\perp=[q/(1+q)^2] M v_\perp^2/2$ and
$T_\perp=-[q/(1+q)^2] GM^2/r_\perp$,
and evaluate their ratio,
\begin{equation}
\label{eqn:rot}
\varrho =\left|\frac{T_\perp}{K_\perp}\right|
=\frac{2 G M}{r_\perp v_\perp^2} =
\frac{c^2}{2}
\frac{\tilde r_{\rm E}\theta_{\rm E}[\tilde r_{\rm E}^{-1}
+(D_{\rm S}\theta_{\rm E})^{-1}]^3}
{d[{\dot d}^2 + (\omega d)^2]}
= \EPST\ \ (D_{\rm S}=10\ \mbox{kpc})
\ .\end{equation}
Here, the error is dominated by the uncertainty in the measurement
of $\omega$.
Since $r_\perp \leq r$ and $v_\perp \leq v$, it follows that
$|T_\perp| \geq |T|$ and $K_\perp \leq K$, and hence
$|T_\perp/K_\perp| \geq |T/K|$, where $|T/K|$ is the ratio of 
the true three-dimensional potential and kinetic energies.
The latter must be greater than unity for a gravitationally bound binary.
This constraint is certainly satisfied by \eb.
Indeed, perhaps it is satisfied too well.
That is, what is the probability of detecting such a large ratio of
projected energies?  To address this question, we numerically
integrate over binary orbital parameters (viewing angles, the orbital
orientation and phase, and the semi-major axis) subject to the constraint
that $r_\perp=5.5\ \mbox{AU}$ ($D_{\rm S}=10\ \mbox{kpc}$),
and at several fixed values of the eccentricity, $e$.
We assume a random ensemble of viewing angles and orbital phases.
The results shown in Figure~\ref{fig:rot} assume a \citet{DM91}
period distribution, but are almost exactly the same if we adopt a flat period
distribution. All of the eccentricities shown in Figure~\ref{fig:rot} are
reasonably consistent with the observed ratio,
although higher eccentricities are favored.

One can also show that
high eccentricities are favored using another line of argument.
First, note that
$|\vect{r}_\perp\times\vect{v}_\perp|=\omega
r_\perp^2$
is the same as the projection of the specific angular momentum of the binary
to the line of sight,
$|(\vect{r}\times\vect{v})\cdot\hat{\vect{n}}|=2\pi a^2(1-e^2)|\cos i|/P$.
Here $a$ and $P$ are the semi-major axis and the orbital period of
the binary, $\hat{\vect{n}}$ is the line-of-sight unit vector, and
$i$ is the inclination angle of the binary.
Combining this result with
Kepler's Third Law, $4\pi^2 a^3=G M P^2$,
one obtains
(here and throughout this section, we assume that $D_{\rm S}=10\ \mbox{kpc}$);
\begin{equation}
\label{eqn:lmin}
a (1-e^2)^2 \cos^2 i =
\frac{(\omega r_\perp^2)^2}{GM}
=\frac{4}{c^2}
\frac{\omega^2 d^4}{\tilde r_{\rm E}\theta_{\rm E}
[\tilde r_{\rm E}^{-1}+(D_{\rm S}\theta_{\rm E})^{-1}]^4}
=\PROJLT
\ .\end{equation}
From the energy conservation,
$(v^2/2)-(GM/r)=-GM/(2a)$,
one may derive a lower limit for $a$;
\begin{equation}
\label{eqn:amin}
a
=\frac{r}{2}\left(1-\frac{rv^2}{2GM}\right)^{-1}
\geq
\frac{r_\perp}{2(1-\varrho^{-1})}
=\AMINT
\ .\end{equation}
The corresponding lower limit for the binary period is
$P\geq 6.22\ \mbox{yr}$ for $D_{\rm S}=10\ \mbox{kpc}$.
Now, we can derive a constraint on the allowed
eccentricity and inclination by dividing equation~(\ref{eqn:lmin})
by equation~(\ref{eqn:amin});
\begin{equation}
(1-e^2) |\cos i| \leq \ECCMINT
\ .\end{equation}
The constraint here essentially arises from the fact that
the fit barely detects projected angular motion $\omega$, while
the formal precision of its measurement corresponds to
$\sim 80\ \mbox{yr}$ orbital period in 1-$\sigma$ level.
For the observed projected separation and the derived binary mass,
this apparent lack of the binary angular motion therefore
naturally leads us to conclude that the binary orbit is either
very close to edge-on or highly eccentric (or both).

\section{Another Look at \mb}

We have found that both parallax and binary orbital motion
are required to explain the deviations from rectilinear motion
exhibited by the light curve of \eb.  In a previous paper
about another event, \mb\ \citep{PL_MB9741}, we had ascribed all deviations
from rectilinear motion to a single cause: projected binary orbital motion.
Could both effects have also been significant in that event as well?
Only detailed modeling can give a full answer to this question. However,
we can give a rough estimate of the size of the projected Einstein ring
$\tilde r_{\rm E}$ that would be required to explain the departures from 
linear motion seen in that event.

Basically what we found in the case of \mb\
was that the light curve in the
neighborhood of the central caustic (near 
HJD $\sim 2450654$)
fixed the lens geometry at that time, and so predicted both the position
of the outlying caustics and the instantaneous
velocity of the source relative to the Einstein ring.
If this instantaneous relative velocity were
maintained, then the source would have missed this outlying caustic by about
$\Delta u_{\rm obs} \sim 0.4$
\citep[fig.~3]{PL_MB9741}.
On the other hand, 
the predicted displacement of a caustic due to parallax is
\begin{mathletters}
\begin{equation}
\Delta\vect{u}_{\rm pred} = 
-\pi_{\rm E} \vect{\mathcal{D}}_{\rm P}
\,;\end{equation}
\begin{equation}
\vect{\mathcal{D}}_{\rm P} =
\vect{\varsigma}_{t_{{\rm cc},1}} - \vect{\varsigma}_{t_{{\rm cc},2}}
- (t_{{\rm cc},1} - t_{{\rm cc},2}) \dot{\vect{\varsigma}}_{t_{{\rm cc},2}}
\ .\end{equation}
\end{mathletters}
We find 
$|\vect{\mathcal{D}}_{\rm P}|=0.072$,
and hence,
\begin{equation}
\pi_{\rm E} = 5.6\
\frac{\Delta u_{\rm pred}}{\Delta u_{\rm obs}}
\ .\end{equation}
That is, to explain  by parallax the order of the effect seen requires
$\tilde r_{\rm E} \sim 0.18\ \mbox{AU}$, 
which (using the measured $\theta_{\rm E}=0.7\ \mbox{mas}$)
would in turn imply a lens mass $M\sim 0.015\ \mbox{M$_\sun$}$,
a lens distance of $D_{\rm L} \la 250\ \mbox{pc}$, and the projected
lens-source relative transverse velocity on the observer plane (at
time $t_{\rm cc,2}$) of only $13\ \mbox{km s$^{-1}$}$. While these values
cannot be strictly ruled out, they are a priori extremely unlikely because
the optical depth to such nearby, low-mass, and slow lenses is extremely
small.  On the other hand, if the lens lies at a distance typical of bulge
lenses $D_{\rm L}\sim 6\ \mbox{kpc}$ and the source is a bulge star, then 
$\tilde r_{\rm E} = D_{\rm rel}\theta_{\rm E} \sim 17\ \mbox{AU}$,
which would imply $\Delta u_{\rm pred}/\Delta u_{\rm obs}\sim 1\%$.
That is, parallax would contribute negligibly to the observed effect.
We therefore conclude that most likely parallax does not play 
a major role in the
interpretation of \mb, but that detailed modeling will be required
to determine to what extent such a role is possible.

What is the physical reason that parallax must be so much stronger
(i.e., $\pi_{\rm E}$ must be so much bigger) to have a significant effect in the
case of \mb\ than for \eb?  Fundamentally, the
former is a close binary, and there is consequently a huge ``lever arm''
between the radial position of the outlying caustic, $u_{\rm c}\sim 1.7$ and
the binary separation, $d\sim 0.5$.  That is $u_{\rm c}/d\sim 3.2$.  This
is almost an order of magnitude larger than for \eb, for which
$u_{\rm c}\sim 0.8$ (Fig.~\ref{fig:geo}), $d\sim 1.9$,
and $u_{\rm c}/d\sim 0.4$.  Consequently,
parallax has to be an order of magnitude larger to reproduce the
effects of the same amount of the projected binary orbital motion.

\section{Summary}
We have presented here results from two seasons of photometric
$I$-band monitoring of the microlensing event \eb,
made by the PLANET collaboration. The light curve exhibits two
peaks which are well explained by a finite source crossing over
a fold-type -- inverse-square-root singularity -- caustic, followed
by a third peak which is due to the source's passage close to a cusp.
We find no geometry involving a rectilinear source-lens relative
trajectory and a static lens that is consistent with the photometry.
However, by incorporating both parallax and binary orbital motion,
we are able to model the observed light curve. In particular,
the detection of the parallax effect is important because it enables
us to derive the microlens mass, $M=\MASS$, unambiguously 
by the combination of the projected Einstein radius
-- measured from the parallax effect -- and the angular Einstein radius
-- inferred from the source angular size and the finite source effect
on the light curve. The source size is measurable from
the magnitude and the color of the source.
The kinematic properties of the lens/source system derived from our model 
together with the lens-source relative parallax measurement
as well as an additional radial velocity measurement
indicate that the event is most likely caused by a (K3) giant star
in the bulge being lensed by a disk binary (M dwarf) system
about $2\ \mbox{kpc}$ from the Sun. Additional information on
the specific density function along the line of sights,
differential reddening across the field,
and a metallicity measurement of the source could further
constrain the source location.

\acknowledgements

We thank the EROS collaboration for providing the initial alert for this
event and for providing photometric data used in the initial real-time
modeling.  We thank the MPS collaboration for providing the secondary
alert that allowed us to densely monitor the critical first caustic crossing.
We are especially grateful to the observatories that support our science 
(Canopus, CTIO, Perth, SAAO) via the generous allocations of time 
that make this work possible. The operation of Canopus Observatory
is in part supported by the financial contribution from
Mr.\ David Warren.
PLANET acknowledges financial support
via award GBE~614-21-009 from de Nederlandse Organisatie voor Wetenschappelijk
Onderzoek (NWO), the Marie Curie Fellowship ERBFMBICT972457
from the European Union (EU), ``coup de pouce 1999'' award from
le Minist\`ere de l'\'Education nationale, de la Recherche et de la
Technologie, D\'epartement Terre-Univers-Environnement, 
a Hubble Fellowship from
the Space Telescope Science Institute (STScI), which is operated by
the Association of Universities for Research in Astronomy (AURA), Inc.,
under NASA contract NAS5-26555, grants AST~97-27520 and AST~95-30619
from the National Science Foundation (NSF), and grants NAG5-7589 and
NAG5-10678 from the National Aeronautics and Space Administration (NASA).

\appendix

\section{Determination of $\vect{\varsigma}$
\label{asec:sun}}
If $\vect{s}$ is the Sun's position vector with respect to the Earth
normalized by $1\ \mbox{AU}$,
then the projection of $\vect{s}$ onto the plane of the sky,
$\vect{\varsigma}$, is
\begin{equation}
\vect{\varsigma} = \vect{s} - (\vect{s}\cdot\hat{\vect{n}})\hat{\vect{n}}
\,,\end{equation}
where $\hat{\vect{n}}$ is the line-of-sight unit vector toward the position
of the event on the sky, while the projection of $\hat{\vect{p}}$,
the unit vector toward the north ecliptic pole (NEP), is given by
$
\tilde{\vect{p}} = 
\hat{\vect{p}} - (\hat{\vect{p}}\cdot\hat{\vect{n}})\hat{\vect{n}}
$.
Then, 
($\varsigma_w,\varsigma_n$),
the ecliptic coordinate components of $\vect{\varsigma}$, are
\begin{mathletters}
\label{eqn:vcrd}
\begin{equation}
\varsigma_w
= \frac{(\tilde{\vect{p}}\times\vect{\varsigma})\cdot\hat{\vect{n}}}
{|\tilde{\vect{p}}|}
= \frac{(\hat{\vect{p}}\times\vect{s})\cdot\hat{\vect{n}}}
{\sqrt{1-(\hat{\vect{p}}\cdot\hat{\vect{n}})^2}}
\,,\end{equation}
\begin{equation}
\varsigma_n
= \frac{\tilde{\vect{p}}\cdot\vect{\varsigma}}{|\tilde{\vect{p}}|}
= 
-\frac{
(\hat{\vect{p}}\cdot\hat{\vect{n}})(\vect{s}\cdot\hat{\vect{n}})}
{\sqrt{1-(\hat{\vect{p}}\cdot\hat{\vect{n}})^2}}
\,,\end{equation}
\end{mathletters}
where we make use of $\hat{\vect{p}}\cdot\vect{s}=0$.
One can choose three-dimensional coordinate axes so that the $x$-axis is the
direction of the vernal equinox, the $z$-axis is the direction to the
NEP, and $\hat{\vect{y}}=\hat{\vect{z}}\times\hat{\vect{x}}$.
Then,
\begin{mathletters}
\label{eqn:crd}
\begin{equation}
\hat{\vect{p}}=(0, 0, 1)
\,;\end{equation}
\begin{equation}
\vect{s}=(r_\earth \cos\lambda_\sun, r_\earth \sin\lambda_\sun, 0)
\,;\end{equation}
\begin{equation}
\hat{\vect{n}}=(\cos\lambda_0\cos\beta_0, 
\sin\lambda_0\cos\beta_0, \sin\beta_0)
\,,\end{equation}
\end{mathletters}
where $r_\earth$ is the distance to the Sun from the Earth in units of AU,
$\lambda_\sun$ is the Sun's ecliptic longitude,
and ($\lambda_0,\beta_0$) are the ecliptic coordinates of the event.
By substituting equations~(\ref{eqn:crd}) into equations~(\ref{eqn:vcrd}),
one finds that
\begin{mathletters}
\begin{equation}
\varsigma_w = - r_\earth \sin(\lambda_\sun - \lambda_0)
\,,\end{equation}
\begin{equation}
\varsigma_n = - r_\earth \cos(\lambda_\sun - \lambda_0) \sin\beta_0
\ .\end{equation}
\end{mathletters}

In general, one must consider the Earth's orbital eccentricity
($\epsilon=1.67\times 10^{-2}$) to calculate $\vect{\varsigma}$
for any given time.
Then,
\begin{mathletters}
\label{eqn:elip}
\begin{equation}
r_\earth = 1 - \epsilon \cos\psi
\,;\ \ \
\lambda_\sun = \xi - \phi_\Upsilon
\,,\end{equation}
\begin{equation}
\psi - \epsilon \sin\psi 
= \Omega t
\,,\end{equation}
\begin{equation}
\sin\xi = \frac{(1-\epsilon^2)^{1/2}\sin\psi}{1-\epsilon\cos\psi}
\,;\ \ \
\cos\xi = \frac{\cos\psi - \epsilon}{1-\epsilon\cos\psi}
\,,\end{equation}
\end{mathletters}
where $\psi$ and $\xi$ are the eccentric and true anomalies of
the Earth,
$\phi_\Upsilon=77\fdg86$ is the true anomaly at the vernal equinox
\citep[March 20, 07$^{\rm h}$35$^{\rm m}$ UT for 2000;][]{almanac},
$t$ is the time elapsed since perihelion,
and $\Omega = 2\pi\ \mbox{yr$^{-1}$}$.
Note that the Earth was at perihelion at January 3,
05$^{\rm h}$ UT for 2000 \citep{almanac}. Although
equations~(\ref{eqn:elip}) cannot be solved for $r_\earth$ and
$\lambda_\sun$ in closed form as functions of $t$,
one can expand in series with respect to $\epsilon$ and
approximate up to the first order (epicycle approximation) so that
\begin{equation}
r_\earth = 1 - \epsilon \cos(\Omega t)
\,;\ \ \
\lambda_\sun = \Omega t - \phi_\Upsilon + 2 \epsilon \sin(\Omega t)
\ .\end{equation}

\section{Microlens Diurnal Parallax
\label{asec:tpar}}
The angular position of a celestial object observed from an
observatory on the surface of the Earth is related to its
geocentric angular position, $\vect{\varphi}_{\rm X}$ by
\begin{equation}
\vect{\varphi}^\prime_{\rm X} =
\vect{\varphi}_{\rm X} - \frac{\vect{\gamma} \mbox{R}_\earth}{D_{\rm X}}
\,,\end{equation}
where
\begin{equation}
\vect{\gamma} = \vect{r} - (\vect{r}\cdot\hat{\vect{n}}) \hat{\vect{n}}
\end{equation}
is the projection of the position vector, $\vect{r}$, of the observatory
with respect to the center of the Earth, onto the plane of the sky and
normalized by the mean radius of the Earth, R$_\earth$; and
$D_{\rm X}$ and $\hat{\vect{n}}$ are
the distance and the line-of-sight vector to the object of interest.
If one observe a microlensing event, the actual dimensionless
lens-source separation vector therefore differs
from $\vect{u}$ (eq.~[\ref{eqn:udef}]) by
\begin{eqnarray}
\label{eqn:tpar}
\vect{u}^\prime
&=&
\frac{\vect{\varphi}^\prime_{\rm S}-\vect{\varphi}^\prime_{\rm L}}
{\theta_{\rm E}}
=
\vect{u} + \frac{\mbox{R}_\earth}{\tilde r_{\rm E}} \vect{\gamma}
\nonumber\\
&=&\vect{\upsilon} + (t-t_{\rm c}) \vect{\mu}_{\rm E} -
\pi_{\rm E}(\vect{\varsigma}-\frac{\mbox{R}_\earth}{1\ \mbox{AU}}\vect{\gamma})
\ .\end{eqnarray}

To find the algebraic expression for ecliptic coordinate
components of $\vect{\gamma}$, we choose
the same coordinate axes as for equations~(\ref{eqn:crd}).
Then, the position vector, $\vect{r}$ is expressed as
\begin{equation}
\vect{r} = (\cos\delta_{\rm g} \cos\tau_\Upsilon,
\cos\delta_{\rm g} \sin\tau_\Upsilon \cos\varepsilon +
\sin\delta_{\rm g} \sin\varepsilon,
\sin\delta_{\rm g} \cos\varepsilon -
\cos\delta_{\rm g} \sin\tau_\Upsilon \sin\varepsilon)
\,,\end{equation}
where $\delta_{\rm g}$ is the geographic latitude of the observatory,
$\tau_\Upsilon$ is the hour angle of the vernal equinox
(i.e.\ the angle of the local sidereal time) at the observation,
and $\varepsilon = 23\fdg44$ is the angle between the direction toward
the north celestial pole and the NEP
(here we also assume that the Earth is a perfect sphere).
Then, following a similar procedure as in Appendix~\ref{asec:sun},
\begin{mathletters}
\begin{equation}
\gamma_w = 
\frac{(\tilde{\vect{p}}\times\vect{\gamma})\cdot\hat{\vect{n}}}
{|\tilde{\vect{p}}|} =
\frac{(\hat{\vect{p}}\times\vect{r})\cdot\hat{\vect{n}}}
{\sqrt{1-(\hat{\vect{p}}\cdot\hat{\vect{n}})^2}}
\,;\end{equation}
\begin{equation}
\gamma_n = \frac{\tilde{\vect{p}}\cdot\vect{\gamma}}{|\tilde{\vect{p}}|} =
\frac{\hat{\vect{p}}\cdot\vect{r}
-(\hat{\vect{p}}\cdot\hat{\vect{n}})(\vect{r}\cdot\hat{\vect{n}})}
{\sqrt{1-(\hat{\vect{p}}\cdot\hat{\vect{n}})^2}}
\,,\end{equation}
\end{mathletters}
\begin{mathletters}
one obtains the ecliptic coordinate components of $\vect{\gamma}$,
\begin{equation}
\gamma_w =
- \sin\delta_{\rm g} \sin\varepsilon \cos\lambda_0
+ \cos\delta_{\rm g} ( \cos\tau_\Upsilon \sin\lambda_0
- \sin\tau_\Upsilon \cos\varepsilon \cos\lambda_0 )
\,;\end{equation}
\begin{equation}
\gamma_n =
\sin\delta_{\rm g}
( \cos\varepsilon \cos\beta_0 - \sin\varepsilon \sin\lambda_0 \sin\beta_0)
- \cos\delta_{\rm g}
[ \cos\tau_\Upsilon \cos\lambda_0 \sin\beta_0 + \sin\tau_\Upsilon
( \sin\varepsilon \cos\beta_0 + \cos\varepsilon \sin\lambda_0 \sin\beta_0) ]
\ .\end{equation}
\end{mathletters}

\section{The Choice of Fit Parameters
\label{asec:fit}}
Judged by the number of fitting parameters alone, \eb\
is by far the most complex event ever analyzed: compared to the runner-up,
\mb\ \citep{MPS_PL, PL_MB9741}, it has two more geometric parameters
and one more limb-darkening parameter. As a direct result,
the path toward choosing a modeling procedure was substantially more tortuous
than usual. We therefore believe that it is important to document this
path, at least in outline, in order to aid in the modeling of future events.

As stated in \S~\ref{sec:par},
the seven standard parameters for binary events are
($d,q,\alpha,u_0,t_{\rm E},t_0,\rho_\ast$).  Immediately following the
first caustic crossing, we fit this crossing to five empirical parameters,
including $t_{\rm cc,1}$ and $\Delta t_1$, the time and half-duration
of the first crossing.  We then changed our choice of parameters
$(t_0,\rho_\ast) \rightarrow (t_{\rm cc,1},\Delta t_1)$
according to the prescription of \citet{PLCS}, effectively
cutting the search space down from seven to five dimensions, and speeding up
the search accordingly.  This permitted us
to accurately predict in real time not only the time, but also the
(4-day) duration of the second crossing which in turn allowed two groups
to obtain large-telescope spectra of the crossing \citep{VLT, Keck}.
This was the first prediction of the duration of a caustic crossing.

Why is the substitution $(t_0,\rho_\ast)\rightarrow
(t_{\rm cc,1},\Delta t_1)$ critical?  Both $t_{\rm cc,1}$ and $\Delta t_1$
are determined from the data with a precision
$\sim 10^{-3}\ \mbox{day} \sim 10^{-5}\ t_{\rm E}$.
Hence, if any of the parameters, $\alpha$, $u_0$, $t_0$, $\rho_\ast$,
are changed \emph{individually} by $10^{-4}$ (subsequent changes of
$t_{\rm cc,1}$ and $\Delta t_1$ by $10^{-4}\ t_{\rm E}$), this will lead to
an increase $\Delta\chi^2 \sim 100$.  As a result, even very modest
movements in parameter space must be carefully choreographed to find a downhill
direction on the $\chi^2$ hypersurface.  By making ($t_{\rm cc,1},\Delta t_1$)
two of the parameters and constraining them to very small steps consistent
with their statistical errors, one in effect automatically enforces this
choreography.

In all the work reported here, we
searched for $\chi^2$ minima at fixed ($d,q$), and
repeated this procedure over a ($d,q$) grid.  We found for this event
(as we have found for others) that regardless of what minimization technique
we apply, if we search ($d,q$) space simultaneously with the other parameters,
then either we do not find the true $\chi^2$ minimum or the search requires
prohibitive amounts of time.

Following the second crossing, we added a linear limb-darkening (LD)
parameter, but otherwise continued with the same parameterization.  We
found that the fitting process was then enormously slowed down because
small changes $\alpha$, $u_0$, or $t_{\rm E}$ led to large changes in
$t_{\rm cc,2}$ and $\Delta t_2$ (the second crossing time and half-duration),
whereas these quantities were directly fixed by the data.  We therefore
changed parameters $(\alpha,u_0,t_{\rm E}) \rightarrow
(t_{\rm cc,2},\Delta t_2,t_{\rm axis})$, where $t_{\rm axis}$ is the time
the source crossed the cusp axis.  Hence, all five of the non-($d,q$)
parameters were fixed more-or-less directly by the data, which greatly
improved the speed of our parameter search.  We thus quickly found the
$\chi^2$ minimum for this (7+1)-parameter --
seven geometry plus one limb-darkening -- space.
(Note that the LD parameter, like the source flux, the background flux,
and the seeing correlation, is determined by linear fitting \emph{after}
each set of other parameters is chosen. Hence, it exacts essentially
zero computational cost.  We therefore track it separately.)

Since the light curve showed systematic residuals of
several percent (compared to daily-averaged photometry errors $\ll 1\%$),
we were compelled to introduce more parameters.  We first added two parallax
parameters; the magnitude $\pi_{\rm E}$ and its relative orientation
with respect to the binary, yielding a (9+1)-parameter fit. Since five of
the nine geometrical parameters remained empirical, this procedure also
converged quickly.  However, while
$\chi^2$ had fallen by several hundred (indicating a very
significant detection of parallax), the problem of systematic residuals was
not qualitatively ameliorated.  This created something of a crisis.  We
realized that further improvements would be possible if we allowed for
binary orbital motion.
Lacking apparent alternatives, we went ahead and introduced
projected binary orbital motion. This led to radical changes in our now
(11+1)-parameterization. Allowing two dimensions of binary motion meant that
both the orientation and the \emph{size} of the binary separation
could change. The latter induced changes in both the size and shape of
the caustic, and so made it essentially impossible to define the
time parameters $\Delta t_{\rm cc}$ in such a way that was at the same
time mathematically consistent and calculable in a reasonable amount of time.
We therefore went back to something very like the original geometric
parameterization but with four additional parameters
($d_{t_0},q,\alpha,u_0,t_{\rm E},t_0,\rho_\ast,
\vect{\pi}_{\rm E},\dot{d},\omega$).
Here the direction of $\vect{\pi}_{\rm E}$ contains the information
on the direction of the binary axis relative to the line of ecliptic latitude 
at time $t_0$, when the source is closest to the binary center of mass
in the Sun's frame of the reference. The binary orbital motion is incorporated
via $\dot{d}$ and $\omega$. The quantities,
$\alpha$, $u_0$, and $t_{\rm E}$ are all given in the frame of the reference
of the Sun as well.

However, while this parameterization has the advantage of mathematical
simplicity, it would have introduced
severe instabilities into the fitting procedure.
At bottom, the problem is the same as the one that led to the substitution
$(t_0,\rho_\ast)\rightarrow (t_{\rm cc,1},\Delta t_1)$ described above, but
substantially more damaging.  This is because the microlens parallax is
a relatively poorly constrained quantity.  Yet, within the framework of this
parameterization, a change of the trial value of $\vect{\pi}_{\rm E}$ of
only 1\% (by itself) would shift $t_{\rm cc,1}$ by
$\sim 0.3\ \mbox{day}$, and so
induce $\Delta\chi^2\sim 1000$.  Thus, to avoid such huge $\chi^2$
jumps, even more careful choreography would have been required.

Instead, we made the following changes to the parameter scheme. First,
we made the reference time, $t_{\rm c}$, be the time of the closest approach
of the source to the cusp, rather than the closest approach to the center
of mass.  Second, we adopted, for the frame of reference, the frame of the
Earth at $t_{\rm c}$ rather than the frame of the Sun.
Our final choice of parameters is
($d_{t_{\rm c}},q,\alpha^\prime,u_{\rm c},t_{\rm E}^\prime,t_{\rm c},
\rho_\ast,\vect{\pi}_{\rm E},\dot{d},\omega$),
where $u_{\rm c}$ is the impact parameter relative to the cusp and
$\alpha^\prime$ and $t_{\rm E}^\prime$ are evaluated 
at $t_{\rm c}$ and in the frame of reference of the Earth
at the time. The change of reference frame
is responsible for the form of the parallax deviation given
in equation~(\ref{eqn:uform}), but from a practical point of view it is very
helpful so that the non-parallax parameters do not change
very much when the parallax is changed: in particular they are similar
to the solution without parallax (i.e.\ $\vect{\pi}_{\rm E} \equiv 0$).
The particular 
choice of the reference time $t_{\rm c}$ is useful because most of the
``action'' of the event happens close to this time, either during the
cusp approach itself, or during the second caustic crossing a week
previously.  Hence, both $t_{\rm c}$ and $u_{\rm c}$
are relatively well fixed by the data, while the angle
$\alpha$ is also relatively well fixed
since it is strongly constrained by the cusp approach and second crossing.
As a consequence, we are able to find relatively robust minima 
for each ($d,q$) grid point in about a single day of computer time,
which is quite adequate to reach a global minimum.

Unfortunately, the (11+1)-parameterization
failed to qualitatively lessen the problem of systematic residuals.
We then recognized that more LD freedom was required, and so added
a square-root LD parameter in addition to the linear one.  This reduced
the systematic residuals to $< 1\%$. As a result,
we fit limb darkening with a two-parameter form so that the final
fit we adopted is an (11+2)-parameterization fit.

\section{Hybrid Statistical Errors
\label{asec:covar}}
As we described in Appendix~\ref{asec:fit},
our $\chi^2$ minimization procedure is
effective at fixed ($d,q$) (with the nine other geometrical parameters
allowed to vary), but does not work when all 11 geometrical parameters
are allowed to vary simultaneously.  We therefore find the global
minimum by evaluating $\chi^2$ over a ($d,q$) grid.  How can the errors,
and more generally the covariances, be determined under these circumstances?

In what follows, the parameters will be collectively represented by a
vector $a_i$ and the indices $i,j$ will be allowed to vary over all $p=11$
parameters.  We assign $a_1=d$ and $a_2=q$, and designate that the indices
$m,n$ will be restricted to these two parameters.  What we seek to
evaluate is $c_{ij} = {\rm cov}(a_i,a_j) \equiv
\langle a_i a_j \rangle - \langle a_i \rangle \langle a_j \rangle$.

First we note that it is straightforward to determine $c_{mn}$: simply
evaluate $\chi^2$ at a series of points on the ($a_1,a_2$) grid, and
fit these to 
$\chi^2 = \chi^2_{\min} + \sum^2_{m,n=1} \hat b_{mn} 
(a_m - a^{\min}_m) (a_n - a^{\min}_n)$.
Then $(\hat c_{mn}) = (\hat b_{mn})^{-1}$ is 
the covariance matrix restricted to
the first two parameters, i.e., $\hat c_{mn} = c_{mn}$.

Next, at fixed $(a_1,a_2)=(a^\dag_1,a^\dag_2)$, 
we evaluate the restricted covariance matrix
of the remaining $p-2=9$ parameters by varying 0, 1, or 2 parameters at
a time and fitting the resulting $\chi^2$ hypersurface to
$\chi^2 = \chi^2_{\rm min} + \sum^{11}_{i,j=3}\tilde b_{ij} 
(a_i - a^{\min}_i) (a_j - a^{\min}_j)$.
Then, using the result derived in the appendix of \citet{GA01},
one can show that
$(\tilde c_{ij}) = (\tilde b_{ij})^{-1}$
is related to the full covariance
matrix by
\begin{equation}
\label{eqn:covar}
\tilde c_{ij} = c_{ij} - \sum_{m,n} \hat b_{mn} c_{mi} c_{nj}
\ \ \
(i,j\neq 1,2)
\,,\end{equation}
and the parameters, $\tilde a_i$, at the constrained minimum are
\begin{equation}
\label{eqn:avar}
\tilde a_i = a^{\min}_i-\sum_{m,n}(a^{\min}_m-a^\dag_m)\hat b_{mn}c_{ni}
\ \ \
(i\neq 1,2)
\ .\end{equation}

Differentiating
equation~(\ref{eqn:avar}) with respect to $a^\dag_m$ yields
\begin{equation}
\frac{\partial \tilde a_i}{\partial a^\dag_m} = \sum_n \hat b_{mn} c_{ni}
\,,\end{equation}
\begin{equation}
\label{eqn:cpart}
c_{mi} = \sum_n \hat c_{mn} \frac{\partial \tilde a_i}{\partial a^\dag_n}
\ .\end{equation}
The partial derivatives can be determined simply
by finding the change in $a_i$ as one steps along one axis of the ($d,q$)
grid.  Since $\hat c_{mn}$ is already known from the first step, above, the
$c_{mi}$ are also known.  Finally, the remaining covariances can be
found by substituting equation~(\ref{eqn:cpart})
into equation~(\ref{eqn:covar}),
\begin{equation}
c_{ij} = \tilde c_{ij} + \sum_{m,n} \hat c_{mn}
{\partial \tilde a_i\over \partial a^\dag_m}
{\partial \tilde a_j\over \partial a^\dag_n}
\ .\end{equation}

\clearpage

\newpage
\begin{deluxetable}{cccccc}
\tablecaption{PLANET $I$-band Photometry of \eb\
\label{tab:dat}}
\tablehead{
\colhead{Telescope}
&\colhead{Error Cut}
&\colhead{Seeing Cut}
&\colhead{Number of Points}
&\colhead{Error Scaling}
&\colhead{Median Seeing}
\\
&\colhead{(mag)}
&\colhead{(arcsec)}
&
&
&\colhead{(arcsec)}
}\startdata
   SAAO&$\leq$ 0.03 \tablenotemark{a}&$\leq$ 2.1&428&1.99&1.41\\
   YALO&$\leq$ 0.03 \tablenotemark{a}&$\leq$ 2.3&424&1.61&1.58\\
Canopus \tablenotemark{b}&    \nodata&   \nodata&333&2.96&2.82\\
  Perth \tablenotemark{b}&    \nodata&$\leq$ 3.1&161&3.63&2.44
\enddata
\tablenotetext{a}{
The formal value reported by DoPHOT.
}\tablenotetext{b}{
The difference imaging analysis result has been used.
}\end{deluxetable}

\begin{deluxetable}{cc}
\tablecaption{Relations between Parameterizations
\label{tab:parm}}
\tablehead{
\colhead{Physical Parameters}&
\colhead{Fit Parameters}
}\startdata
$t_{\rm c}$&$t_{\rm c}$
\\$q$&$q$
\\$\rho_\ast$&$\rho_\ast$
\\$\pi_{\rm E}$&$(\pi^2_{{\rm E},\parallel}+\pi^2_{{\rm E},\perp})^{1/2}$
\\$\vect{d}_{t_{\rm c}}$&
$d_{t_{\rm c}} \pi_{\rm E}^{-1} 
( \pi_{{\rm E},\parallel} \hat{\vect{e}}_w
- \pi_{{\rm E},\perp} \hat{\vect{e}}_n )$
\\$\dot{\vect{d}}_{t_{\rm c}}$&
$\pi_{\rm E}^{-1} [
( \dot d \pi_{{\rm E},\parallel}
+ \omega d_{t_{\rm c}} \pi_{{\rm E},\perp} ) \hat{\vect{e}}_w
+ ( - \dot d \pi_{{\rm E},\perp}
+ \omega d_{t_{\rm c}} \pi_{{\rm E},\perp} ) \hat{\vect{e}}_n ]$
\\$\vect{u}_{t_{\rm c}}$&
$\vect{u}_{\rm cusp} + u_{\rm c} \pi_{\rm E}^{-1} [
( \pi_{{\rm E},\perp} \cos\alpha^\prime
- \pi_{{\rm E},\parallel} \sin\alpha^\prime ) \hat{\vect{e}}_w
+ ( \pi_{{\rm E},\parallel} \cos\alpha^\prime
+ \pi_{{\rm E},\perp} \sin\alpha^\prime ) \hat{\vect{e}}_n ]$
\\$\dot{\vect{u}}_{t_{\rm c}}$&
$t_{\rm E}'^{-1} \pi_{\rm E}^{-1} [
( \pi_{{\rm E},\parallel} \cos\alpha^\prime
+ \pi_{{\rm E},\perp} \sin\alpha^\prime ) \hat{\vect{e}}_w
- ( \pi_{{\rm E},\perp} \cos\alpha^\prime
- \pi_{{\rm E},\parallel} \sin\alpha^\prime ) \hat{\vect{e}}_n ]$
\enddata
\tablecomments{
For simplicity, the reference times, $t_{\rm c}$, for both systems
are chosen to be the same: the time of the closest approach
to the cusp, i.e.\ 
$(\vect{u}_{t_{\rm c}}-\vect{u}_{\rm cusp})\cdot
\dot{\vect{u}}_{t_{\rm c}}=0$.
The additional transformation of the reference time requires
the use of equations~(\ref{eqn:udef}) and (\ref{eqn:uform}).
Figure~\ref{fig:geopar} illustrates the geometry used for
the derivation of the transformation.
The unit vector $\hat{\vect{e}}_n$ points toward the NEP
while $\hat{\vect{e}}_w$ is perpendicular to it and points to the west
(the direction of decreasing ecliptic longitude).
}\end{deluxetable}

\begin{deluxetable}{ccc}
\tablecaption{PLANET Model Parameters for \eb\
\label{tab:fmod}}
\tablehead{\colhead{parameters}&&\colhead{uncertainty\tablenotemark{a}}}
\startdata
$d_{t_{\rm c}}$&$1.928$&$0.004$\\
$q$&$0.7485$&$0.0066$\\
$\alpha^\prime$&$74\fdg18$&$0\fdg41$\\
$u_{\rm c}$&$-5.12\times 10^{-3}$&$3.\times 10^{-5}$\\
$t_{\rm E}^\prime$&$99.8$ days&$1.5$ day\\
$t_{\rm c}$&$1736.944$\,\tablenotemark{b}&$0.005$ day\\
$\rho_\ast$&$4.80\times 10^{-3}$&$4.\times 10^{-5}$\\
$\pi_{{\rm E},\parallel}$&$-0.165$&$0.042$\\
$\pi_{{\rm E},\perp}$&$0.222$&$0.031$\\
$\dot d$&$0.203$ yr$^{-1}$&$0.016$ yr$^{-1}$\\
$\omega$&$0.006$ rad yr$^{-1}$&$0.076$ rad yr$^{-1}$\\
\\
$\mu_{{\rm E},w}$&
$3.83$ yr$^{-1}$&$0.49$ yr$^{-1}$\\
$\mu_{{\rm E},n}$&
$-2.82$ yr$^{-1}$&$0.44$ yr$^{-1}$\\
$\mu_{\rm E}$&$4.76$ yr$^{-1}$&$0.13$ yr$^{-1}$\\
$\alpha_{ec}$\,\tablenotemark{c}&$-36\fdg3$&7\fdg8\\
\\
$\pi_{\rm E}$&$0.277$&$0.008$\\
$\phi$\,\tablenotemark{d}&$-126\fdg5$&$3\fdg6$
\enddata
\tablenotetext{a}
{1-$\sigma$ errorbar. The uncertainties of fit parameters are
determined by fitting $\chi^2$ distribution to a quadratic hypersurface. For
more details, see Appendix~\ref{asec:covar}}
\tablenotetext{b}
{Heliocentric Julian Date $-$ 2450000.}
\tablenotetext{c}
{The angle of $\vect{\mu}_{\rm E}$ with respect to ecliptic west.}
\tablenotetext{d}
{The angle of $\vect{d}_{t_{\rm c}}$ with respect to ecliptic west.}
\end{deluxetable}

\begin{deluxetable}{cr}
\tablecaption{Limb-Darkening Coefficients for \eb\
\label{tab:limb}}
\tablehead{
}\startdata
$\Gamma_I$&$0.452\pm0.075$\\
$\Lambda_I$&$0.011\pm0.137$\\
$\Gamma_I\cos\Phi+\Lambda_I\sin\Phi$\,\tablenotemark{a,} \ \tablenotemark{b}
&$0.207\pm0.156$\\
$\Lambda_I\cos\Phi-\Gamma_I\sin\Phi$\,\tablenotemark{c,} \ \tablenotemark{b}
&$0.402\pm0.003$\\
$c_I$&$0.552\pm0.090$\\
$d_I$&$0.011\pm0.139$\\
$c_I\cos\Psi+d_I\sin\Psi$\,\tablenotemark{a,} \ \tablenotemark{d}
&$0.290\pm0.166$\\
$d_I\cos\Psi-c_I\sin\Psi$\,\tablenotemark{c,} \ \tablenotemark{d}
&$0.470\pm0.003$
\enddata
\tablenotetext{a}
{rotational transformation that maximizes the variance}
\tablenotetext{b}{$\Phi=-61\fdg32$}
\tablenotetext{c}
{rotational transformation that minimizes the variance}
\tablenotetext{d}{$\Psi=-57\fdg14$}
\tablecomments
{The error bars account only for the uncertainty in the
photometric parameters restricted to a fixed lens model, determined by
the linear flux fit.}
\end{deluxetable}

\begin{deluxetable}{cccccc}
\tablecaption{Kinematic Characteristics of the Lens and the Source
\label{tab:kin}}
\tablehead{&
\colhead{location}&
\colhead{$\langle v_y\rangle$\ \tablenotemark{a}}&
\colhead{$\sigma_x$}&
\colhead{$\sigma_y$}&
\colhead{$\sigma_z$}\\&&
\colhead{(km s$^{-1}$)}&
\colhead{(km s$^{-1}$)}&
\colhead{(km s$^{-1}$)}&
\colhead{(km s$^{-1}$)}
}\startdata
$\vect{v}_{\rm L,p}$&disk&$-18$&$38$&$25$&$20$\\
$\vect{v}_{\rm S,p}$&disk&$-11.5$&$31$&$21$&$17$\\
&bulge&\nodata&$100$&$100$&$100$
\enddata
\tablenotetext{a}{
asymmetric drift velocity
}\tablecomments{
The $x$-direction is toward the Galactic center from the LSR,
the $y$-direction is the direction of the Galactic rotation, and
the $z$-direction is toward the north Galactic pole.
The lens is assumed to be an M dwarf while the source is a K3 giant.
The quoted values for the disk components are derived from \citet{BM98}.
}\end{deluxetable}

\end{document}